\PassOptionsToPackage{prologue,dvipsnames}{xcolor} 

\documentclass[acmsmall,screen,nonacm]{acmart}
\setcopyright{cc}
\setcctype{by}

\usepackage{amsmath}
\usepackage{amsthm}
\usepackage{stmaryrd}
\usepackage{mathtools}
\usepackage{enumitem}
\usepackage{pl-syntax}
\usepackage{ifthen}
\usepackage{xspace}
\usepackage{xcolor}
\usepackage{suffix}
\usepackage{centernot}
\usepackage[most]{tcolorbox}
\tcbuselibrary{breakable}
\tcbuselibrary{skins}
\usepackage{tikz}
\usetikzlibrary{decorations.pathreplacing,positioning,calc,arrows,arrows.meta}
\usepackage{tikz-cd}

\tikzset{
  mapsto/.style={{Bar[width=5pt]}-{Stealth[length=5pt,width=4pt]},line width=0.5pt},
  label/.style={font=\small},
  val node/.style={inner sep=0pt,outer sep=0pt},
}

\usepackage{langrules}

\newcounter{numlevels}
\makeatletter
\NewDocumentCommand{\newMLP}{sO{l}mO{0}om}{
  \IfBooleanTF{#1}{
    \IfValueTF{#5}{
      \WithSuffix\newcommand#3*[#4][#5]
    }{
      \WithSuffix\newcommand#3*[#4]
    }
  }{
    \IfValueTF{#5}{
      \newcommand{#3}[#4][#5]
    }{
      \newcommand{#3}[#4]
    }
  }
  {\ifthenelse{\value{numlevels} > 0}{\begin{array}[t]{@{}#2@{}}}{\begin{array}{#2}}\addtocounter{numlevels}{1}#6\addtocounter{numlevels}{-1}\end{array}}
}
\makeatother

\definecolor{darkgreen}{rgb}{0,0.7,0}
\definecolor{navy}{rgb}{0, 0, 0.5}

\definecolor{orangeGold}{rgb}{0.67, 0.33, 0.1}

\colorlet{chorcolor}{RoyalBlue}
\colorlet{localcolor}{PineGreen}
\colorlet{ntwkcolor}{RedOrange}
\colorlet{choicecolor}{orangeGold}

\definecolor{loccolor}{HTML}{b51963}

\makeatletter
\newcommand{\para}{%
  \@startsection{paragraph}{4}{\z@}%
                {-.5\baselineskip plus -2pt minus -.2pt}%
                {-3.5pt}
                {\ACM@NRadjust{\bfseries\@adddotafter}}%
}
\makeatother

\SetRuleLabelLoc{lab}
\SetRuleLabelVCenter
\newcommand{\DefineNtwkRule}[3]{\DefineRule[N#1Rule]{N-#1}{#2}{#3}}
\newcommand{\DefineChorRule}[3]{\DefineRule[C#1Rule]{C-#1}{#2}{#3}}
\newcommand{\DefineKindingRule}[3]{\DefineRule[K#1Rule]{K-#1}{#2}{#3}}
\newcommand{\DefineTypingRule}[3]{\DefineRule[T#1Rule]{T-#1}{#2}{#3}}

\newlength{\vertrulegap}
\newcommand{\programfont}[1]{\ensuremath{\mathsf{#1}}\xspace}
\newcommand{\colrel}[2]{\mathrel{\color{#1}#2}}
\makeatletter
\def\subst@inner#1#2#3\_nil{#1 \mapsto #2\ifx\relax#3\relax\else,\mkern2mu\subst@inner#3\_nil\fi}%
\newcommand{\raw@subst}[3]{
  \mathchoice{#1{\left[#2\subst@inner#3\_nil\right]}}
             {#1[#2\subst@inner#3\_nil]}
             {#1[#2\subst@inner#3\_nil]}
             {#1[#2\subst@inner#3\_nil]}%
}
\newcommand{\subst}[3]{\raw@subst{#1}{}{{#2}{#3}}}
\WithSuffix\newcommand\subst*[2]{\raw@subst{#1}{}{#2}}
\newcommand{\hsubst}[4]{\raw@subst{#1}{#2|}{{#3}{#4}}}
\WithSuffix\newcommand\hsubst*[3]{\raw@subst{#1}{#2|}{#3}}
\makeatother

\makeatletter
\newcommand{\narrowfill@}[5]{%
  $\m@th\thickmuskip0mu\medmuskip\thickmuskip\thinmuskip\thickmuskip
  \relax#5#1\mkern-7mu%
  \cleaders\hbox{$#5\mkern-2mu#2\mkern-2mu$}\hfill
  \mkern-5mu %
  #4%
  \mkern-5mu %
  \cleaders\hbox{$#5\mkern-2mu#2\mkern-2mu$}\hfill
  \mkern-7mu#3$%
}
\newcommand{\nRightarrowfill@}{%
  \narrowfill@\Relbar\Relbar\Rightarrow\neq
}
\newcommand*{\xnRightarrow}[2][]{%
  \ext@arrow 0359\nRightarrowfill@{#1}{#2}%
}
\makeatother

\makeatletter
\DeclareFontEncoding{LS2}{}{\noaccents@}
\DeclareFontSubstitution{LS2}{stix}{m}{n}
\DeclareSymbolFont{stix@largesymbols}{LS2}{stixex}{m}{n}
\SetSymbolFont{stix@largesymbols}{bold}{LS2}{stixex}{b}{n}
\DeclareMathDelimiter{\lBrace}{\mathopen}{stix@largesymbols}{"E8}%
                                         {stix@largesymbols}{"0E}
\DeclareMathDelimiter{\rBrace}{\mathclose}{stix@largesymbols}{"E9}%
                                          {stix@largesymbols}{"0F}
\makeatother

\newcommand{\ty}{\mathrel{:}}
\newcommand{\knd}{\mathrel{::}}
\newcommand{\proves}{\vdash}
\newcommand{\nproves}{\nvdash}
\newcommand{\eproves}{\colrel{localcolor}{\Vdash}}

\newcommand{\seqfun}{\mathrel{\fatsemi}}
\newcommand{\sendsto}{\rightsquigarrow}

\newcommand{\namedlocs}[1]{\operatorname{LN}({#1})}

\newcommand{\fv}[2][]{\operatorname{fv}\if\relax\detokenize{#1}\relax\else_{#1}\fi({#2})}

\newcommand{\lessnd}{\preceq}
\newcommand{\iotasync}{\iota_{\text{sync}}}
\newcommand{\ctrlmerge}{\sqcup}

\newcommand{\val}[1]{\operatorname{Val}({#1})}
\newcommand{\Undef}{\text{undefined}}

\newcommand{\Nec}{\square}
\newcommand{\Pos}{\lozenge}
\newcommand{\Merge}{\sqcup}

\newcommand{\defeq}{=}

\newMLP*[@{}l@{}]{\Newline}[2]{#1 \\ #2}

\newcommand{\say}[1]{\ensuremath{\LocalColor{\lceil}#1\LocalColor{\rfloor}}}
\newcommand{\quoteto}{\ensuremath{\mathrel{\LocalColor{\say{\mathord{\to}}}}}}

\newcommand{\Locations}{\ensuremath{\mathcal{L}}}
\newcommand{\SysLocs}{\ensuremath{\mathfrak{L}}}

\newcommand{\Int}{\programfont{int}}
\newcommand{\Bool}{\programfont{bool}}
\newcommand{\Typ}{\programfont{tyRep}}
\newcommand{\Loc}{\programfont{loc}}
\newcommand{\LocSet}{\programfont{locset}}

\definecolor{orangeGold}{HTML}{AB541A}
\newcommand{\ChoiceCol}[1]{{\color{choicecolor}#1}}
\newcommand{\FontChoice}[1]{\ensuremath{\mathbf{\ChoiceCol{#1}}}\xspace}
\newcommand{\Left}{\FontChoice{L}}
\newcommand{\Right}{\FontChoice{R}}

\newcommand{\ChorCol}[1]{{\color{chorcolor}#1}}
\newcommand{\FontChoreo}[1]{\programfont{\ChorCol{#1}}}
\newcommand{\seq}{\mathrel{\ChorCol{;}}}
\newcommand{\ColSend}{\mathrel{\ChorCol{\sendsto}}}
\newcommand{\NtwkSend}{\mathrel{\CtrlCol{\sendsto}}}
\newcommand{\ChorSend}[1][\ell]{\mathrel{\mathchoice%
  {\raisebox{0.15ex}{$\scriptstyle\ChorCol{\{}#1\ChorCol{\}}$}\mkern-2mu\ChorCol{\sendsto}}
  {\raisebox{0.15ex}{$\scriptstyle\ChorCol{\{}#1\ChorCol{\}}$}\mkern-2mu\ChorCol{\sendsto}}
  {\mkern3mu\raisebox{0.1ex}{$\scriptstyle\ChorCol{\{}#1\ChorCol{\}}$}\mkern-1mu\ChorCol{\sendsto}}
  {\mkern3mu\raisebox{0.05ex}{$\scriptstyle\ChorCol{\{}#1\ChorCol{\}}$}\mkern-1mu\ChorCol{\sendsto}}%
}}
\newcommand{\syncs}[3]{{#1}\ChorCol{[}{#2}\ChorCol{]} \mathrel{\ChorCol{\sendsto}} {#3}}
\newcommand{\LetN}{\FontChoreo{let}}
\newcommand{\In}{\FontChoreo{in}}
\newcommand{\IfN}{\FontChoreo{if}}
\newcommand{\ThenN}{\FontChoreo{then}}
\newcommand{\ElseN}{\FontChoreo{else}}
\newcommand{\FunN}{\FontChoreo{fun}}
\newcommand{\TFunN}{\ChorCol{\Lambda}}
\newcommand{\ChorDef}{\mathrel{\ChorCol{\coloneqq}}}

\newcommand{\InlN}{\FontChoreo{inl}}
\newcommand{\InrN}{\FontChoreo{inr}}
\newcommand{\FstN}{\FontChoreo{fst}}
\newcommand{\SndN}{\FontChoreo{snd}}
\newcommand{\FoldN}{\FontChoreo{fold}}
\newcommand{\UnfoldN}{\FontChoreo{unfold}}

\newcommand{\LetIn}[3]{\LetN~{#1} \ChorDef {#2}~\In~{#3}}
\newMLP*[r@{~}l]{\LetIn}[3]{\LetN & {#1} \ChorDef {#2} \\ \In & {#3}}
\newMLP*{\LetInLine}[3]{\LetN~{#1} \ChorDef {#2}~\In \\ {#3}}
\makeatletter
\def\multilet@inner#1#2#3\_nil{#1 & {}\ChorDef #2\ifx\relax#3\relax\else\\\multilet@inner#3\_nil\fi}%
\newMLP[r@{~}l]{\MultiLet}[2]{\LetN & \begin{array}[t]{@{}l@{}l@{}}\multilet@inner#1\_nil\end{array} \\ \In & {#2}}
\makeatother
\newcommand{\ITE}[4][\rho]{\IfN\if\relax\detokenize{#1}\relax\else_{#1}\fi~{#2}~\ThenN~{#3}~\ElseN~{#4}}
\newMLP*{\ITE}[4][\rho]{\IfN\if\relax\detokenize{#1}\relax\else_{#1}\fi~{#2} \\ \ThenN~{#3} \\ \ElseN~{#4}}
\newcommand{\ITEBase}[3]{\ITE[]{#1}{#2}{#3}}
\newMLP*{\ITEBase}[3]{\ITE*[]{#1}{#2}{#3}}
\newcommand{\Fun}[3]{\mathop{\FunN}{#1}({#2}) \ChorDef {#3}}
\newMLP*{\Fun}[3]{\mathop{\FunN}{#1}({#2}) \ChorDef \\ {#3}}
\newcommand{\TFun}[2]{\TFunN{#1}\ldotp {#2}}

\newcommand{\Inl}[1]{\InlN~{#1}}
\newcommand{\Inr}[1]{\InrN~{#1}}
\newcommand{\CaseN}{\FontChoreo{case}}
\newcommand{\Case}[5]{\CaseN~{#1}~\FontChoreo{of}~(\Inl{#2}\Rightarrow{#3})~(\Inl{#4}\Rightarrow{#5})}
\newMLP*{\Case}[5]{\CaseN~{#1}~\FontChoreo{of} \\ \mid \Inl{#2}\Rightarrow{#3} \\ \mid \Inl{#4}\Rightarrow{#5}}
\newMLP*{\RawCase}[5]{\CaseN~{#1}~\FontChoreo{of} \\ \mid {#2}\Rightarrow{#3} \\ \mid {#4}\Rightarrow #5}
\newcommand{\Fst}[1]{\FstN~{#1}}
\newcommand{\Snd}[1]{\SndN~{#1}}
\newcommand{\Fold}[1]{\FoldN~{#1}}
\newcommand{\Unfold}[1]{\UnfoldN~{#1}}
\newcommand{\None}{\FontChoreo{none}}
\newcommand{\Some}[1]{\FontChoreo{some}~{#1}}

\newcommand{\CtrlCol}[1]{{\color{ntwkcolor}#1}}
\newcommand{\FontCtrl}[1]{\ensuremath{\mathtt{\CtrlCol{#1}}}\xspace}

\newcommand{\CtrlDef}{\mathrel{\CtrlCol{\coloneqq}}}
\newcommand{\AmIN}{\FontCtrl{AmI}}
\newcommand{\AmIInN}{\AmIN\mathord{\CtrlCol{\in}}}

\newcommand{\CtrlThen}{\FontCtrl{then}}
\newcommand{\CtrlElse}{\FontCtrl{else}}
\newcommand{\CtrlLeft}{\Left}
\newcommand{\CtrlRight}{\Right}
\newcommand{\CtrlFunN}{\FontCtrl{fun}}
\newcommand{\CtrlTFunN}{\CtrlCol{\Lambda}}
\newcommand{\CtrlInlN}{\FontCtrl{inl}}
\newcommand{\CtrlInrN}{\FontCtrl{inr}}
\newcommand{\CtrlFstN}{\FontCtrl{fst}}
\newcommand{\CtrlSndN}{\FontCtrl{snd}}
\newcommand{\CtrlFoldN}{\FontCtrl{fold}}
\newcommand{\CtrlUnfoldN}{\FontCtrl{unfold}}
\newcommand{\CtrlVal}[1]{\operatorname{Val}({#1})}

\newcommand{\CtrlLetN}{\FontCtrl{let}}
\newcommand{\CtrlInN}{\FontCtrl{in}}
\newcommand{\CtrlIfN}{\FontCtrl{if}}
\newcommand{\CtrlLetIn}[3]{\CtrlLetN~{#1} \CtrlDef {#2}~\CtrlInN~{#3}}
\newMLP*[r@{~}l]{\CtrlLetIn}[3]{\CtrlLetN & {#1} \CtrlDef {#2} \\ \CtrlInN & {#3}}
\newMLP*{\CtrlLetInLine}[3]{\CtrlLetN~{#1} \CtrlDef {#2}~\CtrlInN \\ {#3}}
\newcommand{\CtrlITE}[3]{\CtrlIfN~{#1}~\CtrlThen~{#2}~\CtrlElse~{#3}}
\newMLP*{\CtrlITE}[3]{\CtrlIfN~{#1} \\ \CtrlThen~{#2} \\ \CtrlElse~{#3}}

\newcommand{\CtrlUnit}{\FontCtrl{()}}
\newcommand{\CtrlNone}{\Undef}
\newcommand{\SendN}{\FontCtrl{send}}
\newcommand{\AllowN}{\FontCtrl{allow}}
\newcommand{\ChoiceN}{\FontCtrl{choice}}
\newcommand{\AllowChoiceN}{\ensuremath{\operatorname{\FontCtrl{allow-choice}}}\xspace}

\newcommand{\Ret}[1]{\FontCtrl{ret}({#1})}
\newcommand{\SendTo}[2]{\SendN~{#1}~\FontCtrl{to}~{#2}}
\newcommand{\RecvFrom}[1]{\FontCtrl{recv}~\FontCtrl{from}~{#1}}

\newcommand{\ChooseFor}[3]{\FontCtrl{choose}~{#1}~\FontCtrl{for}~{#2} \CtrlSeq {#3}}
\newcommand{\AllowChoice}[3]{\AllowN~{#1}~\ChoiceN~(\CtrlLeft \Rightarrow {#2}) ~ (\CtrlRight \Rightarrow {#3})}
\newMLP*{\AllowChoice}[3]{\AllowN~{#1}~\ChoiceN \\ \mid \CtrlLeft \Rightarrow {#2} \\ \mid \CtrlRight \Rightarrow {#3}}
\newcommand{\AllowOneChoice}[3]{\AllowN~{#1}~\ChoiceN~({#2} \Rightarrow {#3})}
\newMLP*{\AllowOneChoice}[3]{\AllowN~{#1}~\ChoiceN \\ \mid {#2} \Rightarrow {#3}}
\newMLP*[@{\mkern2mu}l@{\mkern2mu}]{\AllowChoiceTight}[3]{\AllowN~{#1}~\ChoiceN \\ \mid \CtrlLeft \Rightarrow {#2} \\ \mid \CtrlRight \Rightarrow {#3}}
\newMLP*[@{\mkern2mu}l@{\mkern2mu}]{\AllowOneChoiceTight}[3]{\AllowN~{#1}~\ChoiceN \\ \mid {#2} \Rightarrow {#3}}

\newcommand{\AmI}[3]{\AmIN~{#1}~\CtrlThen~{#2}~\CtrlElse~{#3}}
\newMLP*[r@{~}l]{\AmI}[3]{\AmIN~{#1} & \CtrlThen~{#2} \\ & \CtrlElse~{#3}}
\newcommand{\AmIIn}[3]{\AmIInN~{#1}~\CtrlThen~{#2}~\CtrlElse~{#3}}
\newMLP*[r@{~}l]{\AmIIn}[3]{\AmIInN~{#1} & \CtrlThen~{#2} \\ & \CtrlElse~{#3}}
\newcommand{\CtrlSeq}{\mathrel{\CtrlCol{;}}}

\newcommand{\CtrlInl}[1]{\CtrlInlN~{#1}}
\newcommand{\CtrlInr}[1]{\CtrlInrN~{#1}}
\newcommand{\CtrlCase}[5]{\FontCtrl{case}~{#1}~\FontCtrl{of}~(\CtrlInl{#2}\Rightarrow{#3})~(\CtrlInr{#4}\Rightarrow{#5})}
\newMLP*{\CtrlCase}[5]{\FontCtrl{case}~{#1}~\FontCtrl{of} \\ \mid \CtrlInl{#2}\Rightarrow{#3} \\ \mid \CtrlInr{#4}\Rightarrow{#5}}
\newcommand{\CtrlFst}[1]{\CtrlFstN~{#1}}
\newcommand{\CtrlSnd}[1]{\CtrlSndN~{#1}}
\newcommand{\CtrlFold}[1]{\CtrlFoldN~{#1}}
\newcommand{\CtrlUnfold}[1]{\CtrlUnfoldN~{#1}}

\newcommand{\LocalColor}[1]{{\color{localcolor}#1}}
\newcommand{\LocalLangFont}[1]{\programfont{\color{localcolor}#1}}
\newcommand{\True}{\LocalLangFont{true}}
\newcommand{\False}{\LocalLangFont{false}}
\newcommand{\Loop}{\LocalLangFont{loop}}
\newcommand{\Maybe}[1]{\programfont{maybe}({#1})}
\newcommand{\List}[1]{\programfont{list}({#1})}
\newcommand{\String}{\programfont{string}}
\newcommand{\Nil}{\LocalLangFont{nil}}
\newcommand{\Cons}[2]{\LocalLangFont{cons}({#1},{#2})}
\newcommand{\CtrlFun}[3]{\mathop{\CtrlFunN} {#1}({#2}) \CtrlDef {#3}}
\newMLP*{\CtrlFun}[3]{\mathop{\CtrlFunN} {#1}({#2}) \CtrlDef \\ {#3}}
\newcommand{\CtrlTFun}[2]{\CtrlTFunN{#1}\ldotp {#2}}
\newcommand{\LocalPlus}{\mathrel{\LocalLangFont{+}}}
\newcommand{\LocalTimes}{\mathrel{\LocalLangFont{*}}}
\newcommand{\LocalLess}{\mathrel{\LocalLangFont{<}}}
\newcommand{\LocalLessEq}{\mathrel{\LocalLangFont{\leq}}}
\newcommand{\LocalEq}{\mathrel{\LocalLangFont{=}}}
\newcommand{\LocalITE}[3]{\LocalLangFont{if}~{#1}~\LocalLangFont{then}~{#2}~\LocalLangFont{else}~{#3}}
\newcommand{\LocalCase}[5]{\LocalLangFont{case}~{#1}~\LocalLangFont{of}~({#2} \Rightarrow {#3})~({#4} \Rightarrow {#5})}
\newcommand{\LocalFun}[3]{\mathop{\LocalLangFont{fun}}{#1}({#2}) \mathrel{\LocalColor{\coloneqq}} {#3}}

\newcommand\hole{\ensuremath{[\boldsymbol{\cdot}]}}

\newcommand{\LocFont}[1]{\programfont{\color{loccolor}#1}}
\newcommand{\Client}{\LocFont{C}}
\newcommand{\Worker}{\LocFont{W}}
\newcommand{\Mngr}{\LocFont{M}}
\newcommand{\Alice}{\LocFont{A}}
\newcommand{\Bob}{\LocFont{B}}
\newcommand{\Location}{\LocFont{L}}

\newcommand{\Seller}{\LocFont{S}}
\newcommand{\RunAtWorker}{\programfont{runAtW}}
\newcommand{\WithWorker}{\programfont{runWithWorker}}
\newcommand{\AcquireWorker}{\LocalLangFont{acquireWorker}}
\newcommand{\Pool}{\LocFont{pool}}
\newcommand{\ReleaseWorker}{\LocalLangFont{releaseWorker}}
\newcommand{\Tasks}{\LocalLangFont{tasks}}
\newcommand{\Aggregator}{\programfont{aggregator}}
\newcommand{\Inventory}{\programfont{inventory}}
\newcommand{\Title}{\programfont{title}}
\newcommand{\Budget}{\programfont{budget}}
\newcommand{\Shops}{\LocFont{shops}}
\newcommand{\Lookup}{\programfont{lookup}}
\newcommand{\Price}{\programfont{price}}
\newcommand{\SellerName}{\programfont{seller}}
\newcommand{\Buy}{\programfont{buy}}
\newcommand{\Transact}{\programfont{transact}}

\theoremstyle{theorem}
\newtheorem{thm}{Theorem}

\newtheorem{prop}{Proposition}
\theoremstyle{definition}

\newtheorem{ex}{Example}

\newcommand{\langname}{\ensuremath{\lambda}\textsc{qc}\xspace}

\newcommand\proj[2]{\ensuremath{\left.#1\right|_{#2}}}

\newcommand{\epp}[2]{\mathchoice%
  {\left\llbracket#1\right\rrbracket_{#2}}
  {\llbracket#1\rrbracket_{#2}}
  {\llbracket#1\rrbracket_{#2}}
  {\llbracket#1\rrbracket_{#2}}%
}

\makeatletter
\newlength{\@overset@width}
\newcommand{\raw@step}[3]{\mathrel{%
  \if\relax\detokenize{#1}\relax
    \Longrightarrow
  \else
    \setbox0=\hbox{$\scriptstyle#1$}%
    \setlength{\@overset@width}{\wd0}
    \ifdim\@overset@width<0.75em
      {\overset{#1\mkern4mu}{\Longrightarrow}}{}
    \else
      {\xRightarrow{#1}}{}
    \fi
  \fi%
  \if\relax\detokenize{#2}\relax\else^{#2}\fi%
  \if\relax\detokenize{#3}\relax\else_{#3}\fi%
}}
\newcommand{\raw@reduce}[3]{\mathrel{%
  \if\relax\detokenize{#1}\relax
    \longrightarrow
  \else
    \setbox0=\hbox{$\scriptstyle#1$}%
    \setlength{\@overset@width}{\wd0}
    \ifdim\@overset@width<0.75em
      {\overset{#1\mkern4mu}{\longrightarrow}}{}
    \else
      {\xrightarrow{#1}}{}
    \fi
  \fi%
  \if\relax\detokenize{#2}\relax\else^{#2}\fi%
  \if\relax\detokenize{#3}\relax\else_{#3}\fi%
}}
\newcommand{\step}[1][]{\raw@step{#1}{}{c}}
\newcommand{\stepsn}[2][]{\raw@step{#1}{#2}{c}}
\newcommand{\steps}[1]{\raw@step{#1}{*}{c}}
\newcommand{\stepss}[1][]{\raw@step{#1}{+}{c}}
\newcommand{\localstep}[1][]{\raw@reduce{#1}{}{}}
\newcommand{\localsteps}[1][]{\raw@reduce{#1}{*}{}}
\newcommand{\ctrlstep}[1]{\raw@step{#1}{}{}}
\newcommand{\systemstep}[1][]{\raw@step{#1}{}{S}}
\newcommand{\systemsteps}[1][]{\raw@step{#1}{*}{S}}
\newcommand{\systemstepsn}[2][]{\raw@step{#1}{#2}{S}}
\newcommand{\systemstepss}[1][]{\raw@step{#1}{+}{S}}

\newcommand{\pirstep}{\raw@step{}{}{}}

\newcommand{\raw@nstep}[3]{\mathrel{%
  \if\relax\detokenize{#1}\relax
    {\centernot{\Longrightarrow}}{}
  \else
    {\xnRightarrow{\minwidthbox{0.75em}{\ensuremath{\scriptstyle#1}}}}{}
  \fi%
  \if\relax\detokenize{#2}\relax\else^{#2}\fi%
  \if\relax\detokenize{#3}\relax\else_{#3}\fi%
}}
\newcommand{\nstep}[1]{\raw@nstep{#1}{}{c}}
\newcommand{\nlocalstep}[1][]{\raw@nstep{#1}{}{e}}
\makeatother

\newcommand\localkinded[2]{\ensuremath{{#1} \eproves {#2}}}
\newcommand\localtyped[4]{\localtyped*{#1;#2}{#3}{#4}}
\WithSuffix\newcommand\localtyped*[3]{\ensuremath{{#1} \eproves {#2} \ty {#3}}}

\newcommand\localemptyped[2]{\localtyped*{}{#1}{#2}}
\newcommand\localctxwf[2]{\ensuremath{#1 \eproves #2}}

\newcommand\chorkinded[3]{\ensuremath{{#1} \proves {#2} \knd {#3}}}
\newcommand\chortyped[5]{\chortyped*{#1;#2;#3}{#4}{#5}}
\WithSuffix\newcommand\chortyped*[3]{\ensuremath{{#1} \proves {#2} \ty {#3}}}
\newcommand\choremptyped[2]{\chortyped*{}{#1}{#2}}
\newcommand\chorctxwf[2]{\ensuremath{#1 \proves #2}}

\newcommand{\RDone}[3]{{#1}.({#2} \rightarrow {#3})}
\newcommand{\RArg}[1]{\FontChoreo{Arg}({#1})}
\newcommand{\RFun}[1]{\FontChoreo{Fun}({#1})}
\newcommand{\RPairL}[1]{\FontChoreo{PairL}({#1})}
\newcommand{\RPairR}[1]{\FontChoreo{PairR}({#1})}
\newcommand{\RApp}{\FontChoreo{App}}
\newcommand{\RSendV}[3]{{#1}.{#2} \ColSend {#3}}
\newcommand{\RIfTrue}[1]{\FontChoreo{if}_{#1}~\True}
\newcommand{\RIfFalse}[1]{\FontChoreo{if}_{#1}~\False}
\newcommand{\RTApp}{\FontChoreo{TApp}}
\newcommand{\RLet}[2]{\LetN~{#1} \ChorDef {#2}}
\newcommand{\RUnfoldFold}{\FontChoreo{UnfoldFold}}
\newcommand{\RFstPair}{\FontChoreo{FstPair}}
\newcommand{\RSndPair}{\FontChoreo{SndPair}}
\newcommand{\RCaseInl}{\FontChoreo{CaseInl}}
\newcommand{\RCaseInr}{\FontChoreo{CaseInr}}

\newcommand{\RSendNtwk}[2]{{#1} \NtwkSend {#2}}
\newcommand{\RRecvNtwk}[2]{{#1}.{#2} \NtwkSend}

\newcommand{\cloc}[1]{\operatorname{cloc}({#1})}
\newcommand{\rloc}[1]{\text{rloc}({#1})}

\newcommand{\ctxredex}[1]{{#1}[R]}

\newtcolorbox{mybox}[1][]{
    enhanced,
    breakable,
    title = {#1},
    sharpish corners,
    before skip=\baselineskip,
    leftupper=1.5cm,
    boxrule=0.5pt,
    top=15pt,
    colback=white,
    left=15pt
}

\DefineNtwkRule{Ctx}{
  L \triangleright E_1 \ctrlstep{l} E_2
}{L \triangleright \eta[E_1] \ctrlstep{l} \eta[E_2]}

\DefineNtwkRule{Ret}{
  e_1 \localstep e_2
}{L \triangleright \Ret{e_1} \ctrlstep{\iota} \Ret{e_2}}

\DefineNtwkRule{Seq}{\val{V}}
{L \triangleright V \CtrlSeq E \ctrlstep{\iota} E}

\DefineNtwkRule{App}{
  f = \CtrlFun{F}{X}{E} \\
  \val{V}
  }{L \triangleright f~V \ctrlstep{\iotasync} \subst*{E}{{F}{f}{X}{V}}}

\DefineNtwkRule{TApp}{
  }{L \triangleright (\CtrlTFun{\alpha \knd \kappa}{E})~t \ctrlstep{\iotasync} \subst{E}{\alpha}{t}}

\DefineNtwkRule{UnfoldFold}{\val{V}}
{L \triangleright \CtrlUnfold{(\CtrlFold{V})} \ctrlstep{\iotasync} V}

\DefineNtwkRule{FstPair}{\val{V_1} \\ \val{V_2}}
{L \triangleright \CtrlFst{(V_1,V_2)} \ctrlstep{\iotasync} V_1}

\DefineNtwkRule{SndPair}{\val{V_1} \\ \val{V_2}}
{L \triangleright \CtrlSnd{(V_1,V_2)} \ctrlstep{\iotasync} V_2}

\DefineNtwkRule{CaseInl}{\val{V}}
{L \triangleright \CtrlCase{(\CtrlInl{V})}{X}{E_1}{Y}{E_2} \ctrlstep{\iotasync} \subst{E_1}{X}{V}}

\DefineNtwkRule{CaseInr}{\val{V}}
{L \triangleright \CtrlCase{(\CtrlInr{V})}{X}{E_1}{Y}{E_2} \ctrlstep{\iotasync} \subst{E_2}{X}{V}}

\DefineNtwkRule{Let}{\val{v}
}{L \triangleright \CtrlLetIn{x}{\Ret{v}}{C} \ctrlstep{\iota} \subst{C}{x}{v}}

\DefineNtwkRule{TyLet}{
    \val{\say{t}}
  }{L \triangleright \CtrlLetIn{\alpha \knd \kappa}{\Ret{\say{t}}}{E} \ctrlstep{\iota} \subst*{E}{{\alpha}{t}}}

\DefineNtwkRule{Send}{
  \val{v} \\
  \fv{\rho} = \varnothing
}{L \triangleright \SendTo{\Ret{v}}{\rho} \ctrlstep{\RSendNtwk{v}{\rho \setminus \{L\}}} \Ret{v}}

\DefineNtwkRule{Recv}{
  \val{v} \\
  L' \neq L
}{L \triangleright \RecvFrom{L'} \ctrlstep{\RRecvNtwk{L'}{v}} \Ret{v}}

\DefineNtwkRule{Choose}{
  \fv{\rho} = \varnothing
}{L \triangleright \ChooseFor{d}{\rho}{E} \ctrlstep{\RSendNtwk{d}{\rho \setminus \{L\}}} E}

\newtoggle{NAllowLLinebreak}
\DefineNtwkRule{AllowL}{L' \neq L}{
  L \triangleright \iftoggle{NAllowLLinebreak}{\AllowChoice*{L'}{E_1}{{E_2}_\bot}}{\AllowChoice{L'}{E_1}{{E_2}_\bot}} \ctrlstep{\RRecvNtwk{L'}{\Left}} E_1
}

\DefineNtwkRule{AllowR}{L' \neq L}{L \triangleright \AllowChoice{L'}{{E_1}_\bot}{E_2} \ctrlstep{\RRecvNtwk{L'}{\Right}} E_2}

\DefineNtwkRule{IAmIn}{L \in \rho}
{L \triangleright \AmIIn{\rho}{E_1}{E_2} \ctrlstep{\iota} E_1}

\DefineNtwkRule{IAmNotIn}{L \notin \rho}
{L \triangleright \AmIIn{\rho}{E_1}{E_2} \ctrlstep{\iota} E_2}

\DefineChorRule{Ctx}{
  C_1 \step[R] C_2
}{\eta[C_1] \step[\ctxredex{\eta}] \eta[C_2]}

\DefineChorRule{Done}{
  e_1 \localstep e_2 \\
  \fv{\rho} = \varnothing
}{\rho.e_1 \step[\RDone{\rho}{e_1}{e_2}] \rho.e_2}

\DefineChorRule{App}{
  f = \Fun{F}{X}{C} \\
  \val{V}
}{f~V \step[\RApp] \subst*{C}{{F}{f}{X}{V}}}

\DefineChorRule{TApp}{~}{(\TFun{\alpha \knd \kappa}{C})~t \step[\RTApp] \subst{C}{\alpha}{t}}

\DefineChorRule{UnfoldFold}{
  \val{V}
}{\Unfold{(\Fold{V})} \step[\RUnfoldFold] V}

\DefineChorRule{FstPair}{
  \val{V_1} \\
  \val{V_2}
}{\Fst{(V_1,V_2)} \step[\RFstPair] V_1}

\DefineChorRule{SndPair}{
  \val{V_1} \\
  \val{V_2}
}{\Snd{(V_1,V_2)} \step[\RSndPair] V_2}

\DefineChorRule{CaseInl}{
  \val{V}
}{\Case{(\Inl{V})}{X}{C_1}{Y}{C_2} \step[\RCaseInl] \subst{C_1}{X}{V}}

\DefineChorRule{CaseInr}{
  \val{V}
}{\Case{(\Inr{V})}{X}{C_1}{Y}{C_2} \step[\RCaseInr] \subst{C_2}{X}{V}}

\DefineChorRule{LetV}{
  \val{v} \\
  \fv{\rho} = \varnothing
}{\LetIn{\rho.x \ty t_e}{\rho'.v}{C} \step[\RLet{\rho}{v}] \hsubst{C}{\rho}{x}{v}}

\DefineChorRule{LetI}{
  C_2 \step[R] C_2' \\
  \cloc{C_1} \cap \rloc{R} = \varnothing \\
  \rho \cap \rloc{R} = \varnothing
}{\LetIn{\rho.x \ty t_e}{C_1}{C_2} \step[R] \LetIn{\rho.x \ty t_e}{C_1}{C_2'}}

\newtoggle{TyLetVLinebreak}
\DefineChorRule{TyLetV}{
  \val{\say{t}} \\
  \fv{\rho} = \varnothing
}{
  {\begin{array}{@{}l@{}}
    \LetIn{\rho.\alpha \knd \kappa}{\rho'.\say{t}}{C}
    \iftoggle{TyLetVLinebreak}{ \\ \hspace*{2.5em} }{}
    \step[\RLet{\rho.\alpha}{t}] \subst*{C}{{\alpha}{t}}
  \end{array}}
}

\DefineChorRule{TyLetI}{
  C_2 \step[R] C_2' \\
  \cloc{C_1} \cap \rloc{R} = \varnothing \\
  \rho \cap \rloc{R} = \varnothing
}{\LetIn{\rho.\alpha \knd \kappa}{C_1}{C_2} \step[R] \LetIn{\rho.\alpha \knd \kappa}{C_1}{C_2'}}

\DefineChorRule{SendV}{
  \val{v} \\
  L_1 \in \rho_1 \\
  \fv{\rho_2} = \varnothing
}{\rho_1.v \ChorSend[L_1] \rho_2 \step[\RSendV{L_1}{v}{\rho_2}] (\rho_1 \cup \rho_2).v}

\DefineChorRule{Sync}{
  \fv{\rho} = \varnothing
}{\syncs{L}{d}{\rho} \seq C \step[\RSendV{L}{d}{\rho}] C}

\DefineChorRule{SyncI}{
  C \step[R] C' \\
  \ell \notin \rloc{R} \\
  \rho \cap \rloc{R} = \varnothing
}{\syncs{\ell}{d}{\rho} \seq C \step[R] \syncs{\ell}{d}{\rho} \seq C'}

\DefineChorRule{IfT}{
  \fv{\rho} = \varnothing
}{\ITE[\rho]{\rho.\True}{C_1}{C_2} \step[\RIfTrue{\rho}] C_1}

\DefineChorRule{IfF}{
  \fv{\rho} = \varnothing
}{\ITE[\rho]{\rho.\False}{C_1}{C_2} \step[\RIfFalse{\rho}] C_2}

\DefineChorRule{IfI}{
  C_1 \step[R] C_1' \\
  C_2 \step[R] C_2' \\\\
  \cloc{C} \cap \rloc{R} = \varnothing\\
  \rho \cap \rloc{R} = \varnothing
}{\ITE[\rho]{C}{C_1}{C_2} \step[R] \ITE[\rho]{C}{C_1'}{C_2'}}

\DefineChorRule{AppI}{
  C_2 \step[R] C_2' \\
  \cloc{C_1} \cap \rloc{R} = \varnothing
}{C_1~C_2 \step[R] C_1~C_2'}

\DefineChorRule{PairI}{
  C_2 \step[R] C_2' \\
  \cloc{C_1} \cap \rloc{R} = \varnothing
}{(C_1,C_2) \step[R] (C_1,C_2')}

\DefineKindingRule{Var}{
  \alpha \knd \kappa \in \Gamma
}{\chorkinded{\Gamma}{\alpha}{\kappa}}

\DefineKindingRule{Local}{
  \localkinded{\Gamma}{t_e}
}{\chorkinded{\Gamma}{t_e}{*_e}}

\DefineKindingRule{At}{
  \chorkinded{\Gamma}{t_e}{*_e} \\
  \chorkinded{\Gamma}{\rho}{*_s}
}{\chorkinded{\Gamma}{t_e @ \rho}{*}}

\DefineKindingRule{Arrow}{
  \chorkinded{\Gamma}{\tau_1}{*} \\
  \chorkinded{\Gamma}{\tau_2}{*}
}{\chorkinded{\Gamma}{\tau_1 \to \tau_2}{*}}

\DefineKindingRule{All}{
  \chorkinded{\Gamma, \alpha \knd \kappa}{\tau}{*}
}{\chorkinded{\Gamma}{\forall \alpha \knd \kappa.\tau}{*}}

\DefineKindingRule{Prod}{
  \chorkinded{\Gamma}{\tau_1}{*} \\
  \chorkinded{\Gamma}{\tau_2}{*}
}{\chorkinded{\Gamma}{\tau_1 \times \tau_2}{*}}

\DefineKindingRule{Sum}{
  \chorkinded{\Gamma}{\tau_1}{*} \\
  \chorkinded{\Gamma}{\tau_2}{*}
}{\chorkinded{\Gamma}{\tau_1 + \tau_2}{*}}

\DefineKindingRule{Abs}{
  \chorkinded{\Gamma, \alpha \knd \kappa}{\tau}{*}
}{\chorkinded{\Gamma}{\forall \alpha \knd \kappa.\tau}{*}}

\DefineKindingRule{Rec}{
  \chorkinded{\Gamma, \alpha \knd *}{\tau}{*}
}{\chorkinded{\Gamma}{\mu \alpha.\tau}{*}}

\DefineKindingRule{Loc}{
  L \in \Locations
}{\chorkinded{\Gamma}{L}{*_\ell}}

\DefineKindingRule{Sng}{
  \chorkinded{\Gamma}{\ell}{*_\ell}
}{\chorkinded{\Gamma}{\{\ell\}}{*_s}}

\DefineKindingRule{Union}{
  \chorkinded{\Gamma}{\rho_1}{*_s} \\
  \chorkinded{\Gamma}{\rho_2}{*_s}
}{\chorkinded{\Gamma}{\rho_1 \cup \rho_2}{*_s}}

\DefineRule[WFEmpCtxRule]{WF-EmpCtx}{~}{\chorctxwf{\Gamma}{\cdot}}

\DefineRule[WFAddCtxRule]{WF-AddCtx}{
  X \notin \Delta \\
  \chorctxwf{\Gamma}{\Delta} \\\\
  \chorkinded{\Gamma}{\tau}{*}
}{\chorctxwf{\Gamma}{\Delta, X \ty \tau}}

\DefineRule[WFAddLocalCtxRule]{WF-AddLocalCtx}{
  x \notin \Sigma \\
  \chorctxwf{\Gamma}{\Sigma} \\\\
  \chorkinded{\Gamma}{t_e}{*_e} \\
  \chorkinded{\Gamma}{\rho}{*_s}
}{\chorctxwf{\Gamma}{\Sigma, \rho.x \ty t_e}}

\DefineTypingRule{Var}{
  X \ty \tau \in \Delta \\\\
  \chorctxwf{\Gamma}{\Delta} \\
  \chorctxwf{\Gamma}{\Sigma}
}{\chortyped{\Gamma}{\Delta}{\Sigma}{X}{\tau}}

\DefineTypingRule{Done}{
  \chorkinded{\Gamma}{\rho}{*_s} \\
  \localtyped{\Gamma}{\proj{\Sigma}{\rho}}{e}{t_e} \\\\
  \chorctxwf{\Gamma}{\Delta} \\
  \chorctxwf{\Gamma}{\Sigma}
}{\chortyped{\Gamma}{\Delta}{\Sigma}{\rho.e}{t_e @ \rho}}

\DefineTypingRule{Fun}{
  \chortyped{\Gamma}{\Delta, F \ty \tau_1 \to \tau_2, X \ty \tau_1}{\Sigma}{C}{\tau_2}
}{\chortyped{\Gamma}{\Delta}{\Sigma}{\Fun{F}{X}{C}}{\tau_1 \to \tau_2}}

\DefineTypingRule{App}{
  \chortyped{\Gamma}{\Delta}{\Sigma}{C_1}{\tau_1 \to \tau_2} \\
  \Gamma \vdash C_2 : \tau_1
}{\Gamma \vdash C_1~C_2 : \tau_2}

\DefineTypingRule{Abs}{
  \chortyped{\Gamma, \alpha \knd \kappa}{\Delta}{\Sigma}{C}{\tau}
}{\chortyped{\Gamma}{\Delta}{\Sigma}{\Lambda \alpha \knd \kappa.C}{\forall \alpha \knd \kappa.\tau}}

\DefineTypingRule{TApp}{
  \chortyped{\Gamma}{\Delta}{\Sigma}{C}{\forall \alpha \knd \kappa.\tau} \\
  \chorkinded{\Gamma}{t}{\kappa}
}{\chortyped{\Gamma}{\Delta}{\Sigma}{C~t}{\subst{\tau}{\alpha}{t}}}

\DefineTypingRule{Fold}{
  \chortyped{\Gamma}{\Delta}{\Sigma}{C}{\subst{\tau}{\alpha}{\mu \alpha.\tau}}
}{\chortyped{\Gamma}{\Delta}{\Sigma}{\Fold{C}}{\mu \alpha.\tau}}

\DefineTypingRule{Unfold}{
  \chortyped{\Gamma}{\Delta}{\Sigma}{C}{\mu \alpha.\tau}
}{\chortyped{\Gamma}{\Delta}{\Sigma}{\Unfold{C}}{\subst{\tau}{\alpha}{\mu \alpha.\tau}}}

\DefineTypingRule{Pair}{
  \chortyped{\Gamma}{\Delta}{\Sigma}{C_1}{\tau_1} \\
  \chortyped{\Gamma}{\Delta}{\Sigma}{C_2}{\tau_2}
}{\chortyped{\Gamma}{\Delta}{\Sigma}{(C_1,C_2)}{\tau_1 \times \tau_2}}

\DefineTypingRule{Fst}{
  \chortyped{\Gamma}{\Delta}{\Sigma}{C}{\tau_1 \times \tau_2}
}{\chortyped{\Gamma}{\Delta}{\Sigma}{\Fst{C}}{\tau_1}}

\DefineTypingRule{Snd}{
  \chortyped{\Gamma}{\Delta}{\Sigma}{C}{\tau_1 \times \tau_2}
}{\chortyped{\Gamma}{\Delta}{\Sigma}{\Snd{C}}{\tau_2}}

\DefineTypingRule{Inl}{
  \chortyped{\Gamma}{\Delta}{\Sigma}{C}{\tau_1} \\
  \chorkinded{\Gamma}{\tau_2}{*}
}{\chortyped{\Gamma}{\Delta}{\Sigma}{\Inl{C}}{\tau_1 + \tau_2}}

\DefineTypingRule{Inr}{
  \chortyped{\Gamma}{\Delta}{\Sigma}{C}{\tau_2} \\
  \chorkinded{\Gamma}{\tau_1}{*}
}{\chortyped{\Gamma}{\Delta}{\Sigma}{\Inr{C}}{\tau_1 + \tau_2}}

\DefineTypingRule{Case}{
  \chortyped{\Gamma}{\Delta}{\Sigma}{C}{\tau_1 + \tau_2} \\\\
  \chortyped{\Gamma}{\Delta, X \ty \tau_1}{\Sigma}{C_1}{\tau} \\
  \chortyped{\Gamma}{\Delta, Y \ty \tau_2}{\Sigma}{C_2}{\tau}
}{\chortyped{\Gamma}{\Delta}{\Sigma}{\Case{C}{X}{C_1}{Y}{C_2}}{\tau}}

\DefineTypingRule{LetLocal}{
  \chortyped{\Gamma}{\Delta}{\Sigma}{C_1}{t_e @ \rho_2} \\
  \rho_1 \subseteq \rho_2 \\\\
  \chortyped{\Gamma}{\Delta}{\Sigma, \rho_1.x \ty t_e}{C_2}{\tau}
}{\chortyped{\Gamma}{\Delta}{\Sigma}{\LetIn{\rho_1.x \ty t_e}{C_1}{C_2}}{\tau}}

\DefineTypingRule{LetLoc}{
  \chortyped{\Gamma}{\Delta}{\Sigma}{C_1}{\Loc_{\rho_1} @ \rho_3} \\
  \rho_1 \subseteq \rho_2 \subseteq \rho_3 \\\\
  \chorkinded{\Gamma}{\tau}{*} \\
  \chortyped{\Gamma, \alpha \knd *_\ell}{\Delta}{\Sigma}{C_2}{\tau}
}{\chortyped{\Gamma}{\Delta}{\Sigma}{\LetIn{\rho_2.\alpha \knd *_\ell}{C_1}{C_2}}{\tau}}

\DefineTypingRule{LetLocSet}{
  \chortyped{\Gamma}{\Delta}{\Sigma}{C_1}{\LocSet_{\rho_1} @ \rho_3} \\
  \rho_1 \subseteq \rho_2 \subseteq \rho_3 \\\\
  \chorkinded{\Gamma}{\tau}{*} \\
  \chortyped{\Gamma, \alpha \knd *_s}{\Delta}{\Sigma}{C_2}{\tau}
}{\chortyped{\Gamma}{\Delta}{\Sigma}{\LetIn{\rho_2.\alpha \knd *_s}{C_1}{C_2}}{\tau}}

\DefineTypingRule{LetLocalTy}{
  \chortyped{\Gamma}{\Delta}{\Sigma}{C_1}{\Typ @ \rho_2} \\
  \rho_1 \subseteq \rho_2 \\\\
  \chorkinded{\Gamma}{\tau}{*} \\
  \chortyped{\Gamma, \alpha \knd *_e}{\Delta}{\Sigma}{C_2}{\tau}
}{\chortyped{\Gamma}{\Delta}{\Sigma}{\LetIn{\rho_1.\alpha \knd *_e}{C_1}{C_2}}{\tau}}

\DefineTypingRule{Send}{
  \chortyped{\Gamma}{\Delta}{\Sigma}{C}{t_e @ \rho_1} \\\\
  \ell \in \rho_1 \\
  \chorkinded{\Gamma}{\rho_2}{*_s}
}{\chortyped{\Gamma}{\Delta}{\Sigma}{C \ChorSend \rho_2}{t_e @ (\rho_1 \cup \rho_2)}}

\DefineTypingRule{Sync}{
  \chorkinded{\Gamma}{\ell}{*_\ell} \\
  \chorkinded{\Gamma}{\rho}{*_s} \\\\
  \chortyped{\Gamma}{\Delta}{\Sigma}{C}{\tau}
}{\chortyped{\Gamma}{\Delta}{\Sigma}{\syncs{\ell}{d}{\rho} \seq C}{\tau}}

\DefineTypingRule{If}{
  \chortyped{\Gamma}{\Delta}{\Sigma}{C}{\Bool @ \rho} \\\\
  \chortyped{\Gamma}{\Delta}{\Sigma}{C_1}{\tau} \\
  \chortyped{\Gamma}{\Delta}{\Sigma}{C_2}{\tau}
}{\chortyped{\Gamma}{\Delta}{\Sigma}{\ITE[\rho]{C}{C_1}{C_2}}{\tau}}

\title[Choreographic Quick Changes: First-Class Location (Set) Polymorphism]{Choreographic Quick Changes: \\ First-Class Location (Set) Polymorphism}
\date{}

\author{Ashley Samuelson}
\orcid{0009-0001-8800-2590}
\affiliation{
  \institution{University of Wisconsin--Madison}
  \city{Madison}
  \state{Wisconsin}
  \country{USA}
}
\email{ashley.samuelson@wisc.edu}

\author{Andrew K. Hirsch}
\orcid{0000-0003-2518-614X}
\affiliation{
  \institution{University at Buffalo, SUNY}
  \city{Buffalo}
  \state{New York}
  \country{USA}
}
\email{akhirsch@buffalo.edu}

\author{Ethan Cecchetti}
\orcid{0000-0001-7900-8328}
\affiliation{
  \institution{University of Wisconsin--Madison}
  \city{Madison}
  \state{Wisconsin}
  \country{USA}
}
\email{cecchetti@wisc.edu}

\begin{document}

\begin{abstract}
  Choreographic programming is a promising new paradigm for programming concurrent systems
where a developer writes a single centralized program that compiles to individual programs for each node.
Existing choreographic languages, however, lack critical features integral to modern systems,
like the ability of one node to dynamically compute who should perform a computation and send that decision to others.
This work addresses this gap with \langname, the first typed choreographic language with \emph{first class process names}
and polymorphism over both types and (sets of) locations.
\langname also improves expressive power over previous work by supporting algebraic and recursive data types as well as multiply-located values.
We formalize and mechanically verify our results in Rocq, including the standard choreographic guarantee of deadlock freedom.


\end{abstract}

\maketitle

\section{Introduction}
\label{sec:introduction}
Concurrent programs are integral to many modern software systems, but programming them correctly is notoriously difficult.
Traditionally, each process in the system runs a separate program, but developers must reason about the interactions between these programs and the order in which these events occur.
This complex behavior can easily lead to bugs such as deadlocks, where execution stalls due to nodes with mismatched send and receive expectations forever waiting on each other.

\emph{Choreographic programming}~\citep{Montesi23,Montesi13} is an emerging paradigm that promises to simplify development of correct concurrent systems.
A choreography is a single top-level program that describes the computation performed by every node, including the interactions between them.
A compiler then \emph{projects} programs for individual nodes from this single top-level program.
Choreographies centralize code, putting global control flow in one place and leading to \emph{deadlock-freedom by design},
structurally eliminating a notoriously challenging issue in concurrent code.

The theory of choreographies has advanced rapidly in the last several years
with the addition of higher-order functions~\citep{CruzFilipeGLMP22,HirschG22},
polymorphism over types and processes~\citep{GraversenHM24}, and multiply-located values~\citep{BatesKJSKN25}.
However, none of these results allow for \emph{dynamic computation and sending} of process names in a \emph{type-safe manner}.
The ability to treat host names as first-class values and share them between nodes is critical to many practical applications, such as dynamic load balancers.
The only existing work supporting first-class host names~\citep{SweetDHEHH23} entirely lacks a type system,
and consequently lacks critical type-safety properties.

This paper presents the choreographic Quick Change calculus, \langname, the first choreographic language
to support first-class process names and types.
To understand the value of these features, and choreographic programming in general,
consider a simple cloud computing example where a client~\Client wishes to outsource expensive computation~$F$ on input~$X$.
If~\Client wishes to run~$F$ on a specific (statically known) worker~\Worker, they can do so using the following choreography.
Here $t@\Client$ indicates a value of type~$t$ located at~\Client.
\begin{align*}
  \addtocounter{numlevels}{1}
  & \RunAtWorker : (t \to t') @ \Client \to t@\Client \to t'@\Client \\
  & \RunAtWorker~F~X =
  \def\arraystretch{1.1}
  \MultiLet{{\Worker.f}{F \ColSend \Worker}
            {\Worker.x}{X \ColSend \Worker}}
           {\Worker.(f~x) \ColSend \Client}
  \addtocounter{numlevels}{-1}
\end{align*}
The notation $F \ColSend \Worker$ means that whoever owns $F$ (in this case \Client) should send it to~\Worker.
So in this program, \Client sends the code~($F$) and the input~($X$) to~\Worker, who locally binds them to variables~$f$ and~$x$, respectively.
Then~\Worker computes $f~x$ and sends the result back to~\Client.

Real systems, however, typically include a pool of workers and a load balancer to manage task assignment.
Clients contact this pool manager, who then (a) selects a worker, (b) notifies the worker of their new client, and (c) sends the client the worker's identity.
To implement such a thread pool in a choreography, the client~\Client can ask a pool manager~\Mngr where to run the task,
and~\Mngr must be able to reply with a dynamically chosen identity.
A choreography for such a process might look as follows,
where \AcquireWorker and \ReleaseWorker are local operations by the pool manager~\Mngr to locally track the state of the thread pool and select and return workers to the pool, respectively.
\begin{align*}
  \addtocounter{numlevels}{1}
  & \WithWorker : (t \to t') @ \Client \to t@\Client \to t'@\Client \\
  & \WithWorker~F~X =
  \def\arraystretch{1.1}
  \MultiLet{{W}{\Mngr.\AcquireWorker () \ColSend \{\Client\} \cup \Pool}
            {W.f}{F \ColSend W}
            {W.x}{X \ColSend W}
            {\Client.\mathit{res}}{W.(f~x) \ColSend \Client}}
           {W.\LocalLangFont{``done"} \ColSend \Mngr \seq \Mngr.(\ReleaseWorker~w) \seq \Client.\mathit{res}}
  \addtocounter{numlevels}{-1}
\end{align*}
On the first line of this program, the pool manager~\Mngr selects an idle worker from the pool
and sends this \emph{dynamically computed} value to the client~\Client, where it is bound to the variable~$W$.
Additionally, workers in the~\Pool are all notified about which worker was chosen to prevent situations where a worker is not aware that they have been selected.

As in the earlier \RunAtWorker~function, the client then sends $F$ and $X$ to $W$,
who binds them to~$f$ and~$x$, computes~$f~x$ locally, and sends the result to~\Client.
Finally, the selected worker notifies the manager that they have finished the job, the manager releases the worker back into the thread pool, and the computation finishes by yielding the result.

The first line of this program has very similar syntax to sending and receiving local data, but is critically different:
it sends a location name which is then bound to the variable~$W$ known to all parties in $\{\Client\} \cup \Pool$
and allows the programmer to use it as a location name within its scope.
This is the core feature of \langname: enabling \emph{first-class location names} which can be dynamically computed (and hence quickly changed) at runtime.
While this feature appears simple on its face, it hides several important complexities.
First, the value is known to a set of hosts, not just a single host, making it multiply-located.
Second, location names in choreographies are generally part of the language of \emph{types}, not values.
To support first-class location names without the need for dependent types,
we carefully separate locations from other values in the \langname type system, while still allowing interaction between the two.
These key insights allow us to intermingle dynamically chosen host names with process polymorphism, such as in the example above.

The main contributions of this work are as follows.
\begin{itemize}[leftmargin=*]
  \item We generalize the constraints on the local (message) language of \emph{Pirouette}~\citep{HirschG22} to allow for polymorphism, vastly increasing the expressivity of local computations (Section \ref{sec:system_model}).
  \item We present \langname, the first typed choreographic programming language with the ability to send and receive first-class location names and types as messages.
  Our language includes polymorphism over types, processes, and sets of processes, algebraic and recursive data types, and multiply-located values.
  We formulate a sound, System F-like type system for our language without relying on dependent types or an operational semantics for types (Section \ref{sec:language}).
  \item We define a network language (Section \ref{sec:network-lang}) which serves as the compilation target for our endpoint projection procedure.
  We show that compilation is complete, and is sound when all executed local computations are terminating.
  This result, along with the soundness of our type system, allows us to prove that projected systems are always deadlock-free (Section \ref{sec:endpoint-projection}).
  \item We formalize and verify all results in the Rocq Prover (formerly Coq).
    This is the first mechanized formalization both of process (set) polymorphism and of multiply-located values (Section \ref{sec:rocq-dev}).
\end{itemize}


\section{Background}
\label{sec:background}
To understand the contributions of this work, it is helpful to understand some background on choreographies with higher-order functions, process polymorphism, and multiply-located values.

\subsection{Pirouette}
\label{sec:bg-pirouette}
We generalize and build on Pirouette~\citep{HirschG22}, a higher-order choreographic programming language.
As with other choreographies, Pirouette uses a located syntax inspired by the ``Alice and Bob'' syntax of cryptographic protocols.
To specify that~\Alice should (locally) compute $2 \LocalPlus 3$ and send the result to \Bob, one would write $\Alice.(2 \LocalPlus 3) \ColSend \Bob$.
For clarity, we use \programfont{sans\text{-}serif} for these source-level operations and color
choreographic operations in $\FontChoreo{blue}$,
local operations in $\LocalLangFont{green}$,
and location constants in $\LocFont{red}$.

One core aspect of Pirouette's design that we inherit is a clear separation between choreographic operations and local operations.
The choreographic operations are fixed by Pirouette, and are agnostic to the local language. 
The local operations can be specified in nearly any language whose values---possibly including functions---can be communicated between nodes via message-passing.
The local language need only have an operational semantics, a type system, and a type that can act as a boolean
(every value of that type can be used as \True or \False).

To connect the choreographic and local languages, Pirouette offers a form of let-expression that binds local variables to the output of a choreography.
For instance, the result of $\Alice.(2 \LocalPlus 3) \ColSend \Bob$ is a choreographic value located at \Bob,
so using it in \Bob's local computation requires binding a \emph{local} variable with the output of a \emph{choreographic} computation.
The following program demonstrates how~\Bob can perform this binding and use the result locally as the variable~$x$.
\[
  \LetIn{\Bob.x}{(\Alice.(2 \LocalPlus 3) \ColSend \Bob)}{\Bob.(4 \LocalTimes x)}
\]

Pirouette also supports conditionals $\ITEBase{C}{C_1}{C_2}$ where the choreography itself branches based on the result of~$C$, which should yield a boolean at some location~$\ell$.
However, only~$\ell$ can access the branch condition,
so others with differing behavior in the~\textbf{\ChoiceCol{L}}(eft) and~\textbf{\ChoiceCol{R}}(ight)
branches cannot proceed without knowing which to take.
The location~$\ell$~can communicate the choice~$d \in \{\Left,\Right\}$ of branch to another location~$\ell'$ using a \emph{selection statement} $\syncs{\ell}{d}{\ell'} \seq C$.

At the choreographic level, Pirouette is simply typed, containing only base types and function types.
The base types are of the form~$t @ \ell$, and describe a value of local type~$t$---any type from the local language's type system---located at~$\ell$.
This mixing of the local and choreographic type systems and the local variables bound by let-expressions, shown above, forces the type system to separately track both choreographic and local variables.

Pirouette defines two separate semantics:
one directly at the choreographic level to allow developers to more easily reason about the behavior of the choreography,
and one using a translation called \emph{endpoint projection}~(EPP) that defines how to compile a choreography to separate programs for each location.
Despite operating directly on a choreography, the first semantics also captures concurrent interleavings
by allowing out-of-order execution as long as it does not reorder the operations for any individual location.
For example, consider the program shown below.
\[
  \Alice.(2 \LocalPlus 3) \ColSend \Bob \seq \Client.(5 \LocalTimes 4) \ColSend \Bob
\]
The local computations for~\Alice and~\Client involve disjoint parties, so they can execute in either order.
The sends, however, both involve~\Bob and must occur in the order specified:
\Bob~must receive~5 from~\Alice before receiving~20 from~\Client.

EPP defines how to compile a choreography into separate programs for each location that include only the operations that location needs to run.
The location being projected to is denoted by a subscript to the projection operator, as in $\epp{C}{\Alice}$ and $\epp{C}{\Bob}$.
For instance, sending a value from~\Alice to~\Bob requires~\Alice to compute the value and send it, and requires~\Bob to receive from \Alice:
\begin{mathpar}
  \epp{\Alice.(2 \LocalPlus 3) \ColSend \Bob}{\Alice} = \SendTo{\Ret{2 \LocalPlus 3}}{\Bob}
  \and
  \epp{\Alice.(2 \LocalPlus 3) \ColSend \Bob}{\Bob} = \RecvFrom{\Alice}
\end{mathpar}
We write operations of our target (network) language in $\FontCtrl{orange~teletype}$ font.

\citet{HirschG22} prove that running these projected programs in parallel produces the same result as the first semantics,
meaning that developers can safely reason using the top-level semantics and, critically, any projected choreography is deadlock-free,
meaning it cannot get stuck with locations waiting to receive messages from each other (or for any other reason).

\subsection{Process Polymorphism}
While this work is the first to support \emph{first-class} locations in a typed choreography,
PolyChor$\lambda$~\citep{GraversenHM24} introduced the notion of process polymorphism,
providing a way to abstract over the participants in a choreography.
PolyChor$\lambda$ introduces a \emph{process abstraction}~$\TFun{\ell}{C}$, akin to a classic type abstraction,
that binds variable~$\ell$ representing a process name in choreography~$C$.
In our syntax, a programmer could write the following process function in which~\Alice computes a value and sends it to a yet to be determined recipient~$\ell$.
\[
  F = \TFun{\ell}{\Alice.(2 \LocalPlus 3) \ColSend \ell}
\]
In PolyChor$\lambda$, the only way to use a process abstraction is to apply it to a location (e.g., $F~\Bob$ or $F~\Client$)
or a location variable created by another process abstraction  (e.g., $F~\ell$).
Location names are not first-class values that can be computed locally or sent between parties.

Despite process names being statically resolvable in PolyChor$\lambda$, the projection of a process abstraction requires locations to behave differently depending on what the variable~$\ell$ resolves to \emph{at run time}.
This is accomplished using an ``AmI'' construct in the compiled language that produces a local branch and reflects the fact that each process should know its own identity.
Specifically, the program $\AmI{\Alice}{E_1}{E_2}$ will execute~$E_1$ when running at~\Alice and~$E_2$ when running elsewhere.
EPP can then project a process abstraction to an $\AmIN$ statement where each branch is a different projection of~$C$:
$\CtrlThen$ replaces~$\ell$ with the current process, while $\CtrlElse$ assumes~$\ell$ has resolved to a different name.
That is,
\[
  \epp{\TFun{\ell}{C}}{\Alice} = \CtrlTFun{\ell}{\AmI{\ell}{\epp{\subst{C}{\ell}{\Alice}}{\Alice}}{\epp{C}{\Alice}}}.
\]

A feature of many polymorphic languages which is notably missing from PolyChor$\lambda$ is \emph{recursive types}.
The addition of recursive types to this language would be challenging because their type system---based on System $F_\omega$---includes an operational semantics, and as a result their endpoint projection procedure requires types which appear in the choreography to be fully-reduced type values.
Implementing this feature would require either that endpoint projection handle
choreographies in which nonterminating types may appear, or that recursive
types are limited in order to force type computations to converge.
Both of these options have significant drawbacks.

\subsection{Multiply-Located Values}
An alternative to the selection messages Pirouette and similar choreographies
use in choreographic conditionals is \emph{multiply-located values}~\citep{BatesKJSKN25,SweetDHEHH23}.
These generalize the notion of local values at one location to local values at a set of locations.
This feature recognizes that, after sending a value, all parties involved know that value,
allowing the choreography to guarantee that the locations agree.

For instance, the choreography $\{\Alice,\Bob\}.(2 \LocalLess 3) \ChorSend[\Alice] \Client$
specifies that~\Alice and~\Bob should \emph{both} compute $2 \LocalLess 3$ and~\Alice should send the result to~\Client.
Notably, the result of this computation is the multiply-located boolean value $\{\Alice, \Bob, \Client\}.\True$,
meaning a condition could branch on the result without requiring further synchronization between these three parties.


\section{System Model}
\label{sec:system_model}
Before introducing the choreographic constructs in \langname, we specify the assumptions on the setting in which it operates.
As is standard in the choreography literature, we assume a fixed set of locations~\Locations, each with a unique name.
Names are taken to be opaque identifiers associated with an underlying node, process, thread, etc.

\label{sec:local-programs}
Like Pirouette~\citep{HirschG22}, \langname abstracts over the language for local computation, requiring only that it satisfy a small set of rules.
To support type and location polymorphism and multiply-located values, \langname adds a few assumptions beyond prior work,
but it still supports numerous languages.

The local language must have a syntax that specifies a set of values, a small-step operational semantics, and a type system.
We write $e_1 \localstep e_2$ to denote that a local term~$e_1$ steps to~$e_2$ in the local language's semantics.
The semantics must satisfy two properties:
\begin{enumerate}[topsep=3pt,itemsep=2pt]
  \item\label{prop:li:values} Values cannot step. That is, if $\val{v}$, then there is no~$e$ such that $v \localstep e$.
  \item\label{prop:li:confluence} Local steps satisfy the \emph{diamond property}.
    That is, if $e_1 \localstep e_2$ and $e_1 \localstep e_3$,
    then either $e_2 = e_3$ or there is some~$e_4$ such that $e_2 \localstep e_4$ and $e_3 \localstep e_4$.
\end{enumerate}
Property~(\ref{prop:li:values}) is taken from Pirouette,
while property~(\ref{prop:li:confluence}) ensures multiply-located computations all produce the same result.
Property~(\ref{prop:li:confluence}) may appear restrictive, but many pure functional languages and all deterministic languages enjoy it.
We believe it can be weakened to the more general confluence statement that,
if $e_1 \localsteps e_2$ and $e_1 \localsteps e_3$ then $e_2 \localsteps e_4$ and $e_3 \localsteps e_4$ for some~$e_4$,
but we leave the challenge of mechanically verifying this conjecture to future work.

\subsection{Local Kinding and Type System}
\label{sec:local-types}
As \langname is constructed generically over the local language, \langname's type system---and the guarantees it provides---depend on the local language specifying a type system.
In particular, we require the local language to specify both a type system and a kinding system.
For simplicity, we show only a single local kind~($*_e$) here, but the results generalize to an arbitrary non-empty set of local kinds, which our Rocq formalization allows.
The local type system can be polymorphic, but this is not required---it may instead be a simple type system.

\begin{figure}
  \begin{ruleset}
    \Rule{LK-Var}{\alpha \in \Gamma}
    {\localkinded{\Gamma}{\alpha}}
    \and
    \Rule{LK-TSubst}{\localkinded{\Gamma}{t_1} \\
      \localkinded{\Gamma, \alpha}{t_2}}
    {\localkinded{\Gamma}{\subst{t_2}{\alpha}{t_1}}}
    \and
    \Rule{LT-WF}{\localtyped{\Gamma}{\Sigma}{e}{t}}
    {\localctxwf{\Gamma}{\Sigma}}
    \and
    \Rule{LT-Kind}{\localtyped{\Gamma}{\Sigma}{e}{t}}
    {\localkinded{\Gamma}{t}}
    \\
    \Rule{LT-Var}{\localctxwf{\Gamma}{\Sigma} \\
      x \ty t \in \Sigma}
    {\localtyped{\Gamma}{\Sigma}{x}{t}}
    \and
    \Rule{LT-Subst}{\localtyped{\Gamma}{\Sigma}{e_1}{t_1} \\
      \localtyped{\Gamma}{\Sigma, x \ty t_1}{e_2}{t_2}}
    {\localtyped{\Gamma}{\Sigma}{\subst{e_2}{x}{e_1}}{t_2}}
    \and
    \Rule{LT-TSubst}{\localkinded{\Gamma}{t_1} \\
      \localtyped{\Gamma, \alpha}{\Sigma}{e_2}{t_2}}
    {\localtyped{\Gamma}{\subst{\Sigma}{\alpha}{t_1}}{\subst{e_2}{\alpha}{t_1}}{\subst{t_2}{\alpha}{t_1}}}
  \end{ruleset}
  \caption{Required-admissible kinding and typing rules for the local language.}
  \label{fig:local-rules}
\end{figure}

The local kinding and type systems are specified by a type well-formedness judgment~\localkinded{\Gamma}{t} and an expression typing judgment~\localtyped{\Gamma}{\Sigma}{e}{t}, respectively.
The double-vertical turnstile~$\eproves$ indicates the local kind and type systems.
In each judgment~$\Gamma$ is a kinding context, and~$\Sigma$ is a typing context.
All we require is that the rules in Figure~\ref{fig:local-rules} be \emph{admissible}.
Admissibility is a standard notion in proof theory~\citep{Takeuti87}
meaning a rule must be \emph{provable} without requiring that it (necessarily) be an axiom of the system.
For instance, \ruleref{LT-Kind} requires the type of any expression be well-kinded,
and \ruleref{LT-Subst} (resp.~\ruleref{LT-TSubst}) requires expression (resp.~type) substitution to be type-preserving.
In most standard polymorphic type systems,
including the local languages presented below in Section~\ref{sec:loc-lang-examples},
these rules are not axioms, but they are provable and thus admissible.

As in Pirouette, the local type system must include a type, which we write as~\Bool,
where every value can be interpreted as either \True or~\False.
These are required to define the behavior of conditionals, which branch on local values.
We also require three more types:
$\Loc_\rho$, $\LocSet_\rho$, and \Typ,
defining the first-class local \emph{representations} of location names, sets of locations, and local types, respectively.
The defining property of representations is that we can convert values of type $\Loc_\rho$ to location names in~\Locations,
values of type $\LocSet_\rho$ to \emph{sets} of location names,
and values of type \Typ to well-formed types in the local language.
For example, one could represent location names using integers, mapping 0 to~\Alice, 1 to~\Bob, and all other integers to~\Client.
We write representations of a type, location name, or set of location names as~$\say{t}$, ~$\say{L}$, or~$\say{\{L_1,\ldots,L_n\}}$ respectively, to disambiguate from the resolved type~$t$, location~$L$, or set of locations $\{L_1,\ldots,L_n\}$.
In the above example, $\say{\Alice}$~is syntactic sugar for 0, while~\Alice refers to the \emph{actual} location $\Alice \in \Locations$.
A location need not have a unique representative, but each value can only represent one location.
There is also no requirement that all possible booleans, location names, etc. be present,
or even that the local types be inhabited, but uninhabited types will render choreographic features unavailable.

The subscript~$\rho$ in the type~$\Loc_\rho$ is an upper bound on the set of locations to which an expression of that type may resolve.
For instance, both $\say{\Alice}$ and $\LocalITE{e}{\say{\Alice}}{\say{\Bob}}$ can have type $\Loc_{\{\Alice,\Bob\}}$, but~$\say{\Client}$ cannot.
For~$\LocSet_\rho$, the annotation must be a superset of any value of this type (e.g., $\say{\{\Alice,\Bob\}} : \LocSet_{\{\Alice,\Bob,\Client\}}$).
The precision of~$\rho$ does not explicitly affect the operational semantics of our language,
but an imprecise annotation may require a choreography to add unnecessary communication to remain well-typed (see Section~\ref{sec:type-system}).
In practice, this static upper bound could be computed in a multitude of ways, such as refinement types~\citep{Freeman91} or a separate static analysis.
In this work, we assume for simplicity that the annotations are given directly by the local type system, and leave computing them precisely to future work.

Lastly, the local type system must provide standard progress and preservation guarantees:
\begin{itemize}[itemsep=2pt,topsep=2pt]
  \item \textit{Local Progress:}\hspace{1ex}%
    If $\localemptyped{e}{t}$ then either~$e$ is a value, or there is some~$e'$ such that $e \localstep e'$.
  \item \textit{Local Preservation:}\hspace{1ex}%
    If $\localtyped{\Gamma}{\Sigma}{e}{t}$ and $e \localstep e'$, then $\localtyped{\Gamma}{\Sigma}{e'}{t}$.
\end{itemize}

\subsubsection{Example Local Languages}\label{sec:loc-lang-examples}
Many simply-typed and polymorphic $\lambda$-calculi satisfy our requirements with very minor modification.
To show that our requirements are both reasonable and satisfiable we present two examples here, with the full details contained in our formalization.

\begin{ex}[Simply-Typed $\lambda$-Calculus]\label{ex:st-lambda}
  The simply-typed, call-by-value $\lambda$-calculus with primitive integers satisfies our requirements.
  Specifically, integers can represent both booleans (zero is~$\False$ while all non-zero values are~$\True$) and location names.
  Having many representations of~\True is not a concern, as each representation remains unambiguous.
  The space of both location sets and types, $\LocSet_\rho$~and~\Typ, respectively, can both be the empty type,
  as not all representations are required to exist.
  This choice will render first-class  location sets and local types unavailable at the choreographic level,
  but does not otherwise interfere with the availability of choreographic type-, location-, or location-set--polymorphism.
\end{ex}

\begin{ex}[System F]\label{ex:system-f}
  System F with primitive booleans, integers representing locations, and lists of integers representing location sets satisfies our requirements.
  Like using integers for~\True and~\False, having multiple ordered lists represent the same set of locations is not a concern.
  We can also include recursive functions, as there is no requirement that local expressions terminate.

  To represent a non-trivial set of local types, we can define~\Typ to be a primitive data type with constructors \LocalColor{\say{\Int}}, \LocalColor{\say{\Bool}}, and $e_1 \quoteto e_2$.
  Since the representations are expressions not types, \Typ~behaves identically to the other primitive types of the language.
  The syntax of the resulting example language is shown below.
  \begin{syntax}
    \category[Types]{t} \alternative{\alpha} \alternative{\Loc_\rho} \alternative{\LocSet_\rho} \alternative{\Int} \alternative{\Bool} \alternative{\List{t}} \alternative{\Typ} \alternative{t_1 \to t_2} \alternative{\forall \alpha.t}
    \category[Expressions]{e} \alternative{x} \alternative{\LocalFun{f}{x \ty t}{e}} \alternative{e_1~e_2} \alternative{\LocalColor{\Lambda} \alpha.e} \alternative{e~t}\\
    \alternative{n \in \mathbb{Z}} \alternative{e_1 \LocalPlus e_2} \alternative{e_1 \LocalEq e_2} \alternative{e_1 \LocalLess e_2}\\
    \alternative{\True} \alternative{\False} \alternative{\LocalITE{e}{e_1}{e_2}}\\
    \alternative{\Nil} \alternative{\Cons{e_1}{e_2}} \alternative{\LocalCase{e}{\Nil}{e_1}{\Cons{x}{y}}{e_2}} \\
    \alternative{\LocalColor{\say{\Int}}} \alternative{\LocalColor{\say{\Bool}}} \alternative{e_1 \quoteto e_2}
  \end{syntax}

  As an example of how the type system is used to provide a static upper-bound on representations of locations, two of the typing rules for the~$\Loc_\rho$ type are given below.
  \begin{mathpar}
    \infer{n \in \rho \subseteq \mathbb{Z}\\
    \localctxwf{\Gamma}{\Sigma}}
      {\localtyped{\Gamma}{\Sigma}{n}{\Loc_\rho}}
    \and
    \infer{\localtyped{\Gamma}{\Sigma}{e}{\Bool}\\
      \localtyped{\Gamma}{\Sigma}{e_1}{\Loc_{\rho}}\\
      \localtyped{\Gamma}{\Sigma}{e_2}{\Loc_{\rho}}}
      {\localtyped{\Gamma}{\Sigma}{\LocalITE{e}{e_1}{e_2}}{\Loc_{\rho}}}
  \end{mathpar}
\end{ex}


\section[The Lambda-QC Language]{The \langname Language}
\label{sec:language}
We now present the Quick Change Choreographic calculus~ (\langname),
a polymorphic $\lambda$-calculus for choreographies
that supports communication of dynamically generated (sets of) location names and types,
as well as multiply-located values and algebraic and recursive data types,
while retaining the traditional guarantee of deadlock freedom.

\subsection[Lambda-QC Syntax]{\langname Syntax}
\label{sec:syntax}

\begin{figure}[b]
  \begin{syntax}
    \category[Selection Labels]{d}
    \alternative{\Left}
    \alternative{\Right}

    \category[Choreographies]{C}
    \alternative{X} \alternative{\rho.e} \alternative{\Fun{F}{X}{C}} \alternative{C_1~C_2} \alternative{\TFun{\alpha \knd \kappa}{C}} \alternative{C~t}\\
    \alternative{\Fold{C}} \alternative{\Unfold{C}} \alternative{(C_1,C_2)} \alternative{\Fst{C}} \alternative{\Snd{C}} \\
    \alternative{\Inl{C}} \alternative{\Inr{C}} \alternative{\Case{C}{X}{C_1}{Y}{C_2}}\\
    \alternative{C \ChorSend[\ell] \rho} \alternative{\syncs{\ell}{d}{\rho} \seq C}
    \alternative{\ITE{C}{C_1}{C_2}}\\
    \alternative{\LetIn{\rho.x \ty t}{C_1}{C_2}} \alternative{\LetIn{\rho.\alpha \knd \kappa}{C_1}{C_2}}\\
  \end{syntax}

  \caption{Syntax of Choreographies}
  \label{fig:abstract-syntax}
\end{figure}

Figure~\ref{fig:abstract-syntax} presents the full syntax of \langname.
To visually differentiate classes of variables, we write choreographic program variables in uppercase Roman characters ($X, Y, F, \dotsc$),
local program variables in lowercase Roman characters ($x, y, f, \dotsc$),
and type, location, and location set variables in lowercase Greek characters ($\alpha, \beta, \dotsc$).
The metavariable~$\ell$ denotes a location, $\rho$~a~set of locations, $\tau$~a~choreographic type, $t$~a~local type, and $\kappa$~a~kind.

Much of the \langname syntax consists of standard algebraic and recursive datatypes lifted to choreographies, but there are a few forms of note.
First, the term~$\rho.e$ specifies that the set of locations~$\rho$ should run local program~$e$.
Note that~$\rho$ must be non-empty, as an empty set of locations performing a computation would be meaningless.
We use the shorthand~$\ell.e$ to mean~$\{\ell\}.e$.
Local programs can use local variables bound in the scope of either the choreography or the local program itself.
For these variables, every location has a separate namespace, so~$\Alice.x$ denotes variable~$x$ in the namespace of~\Alice.
Each location's namespace is separate, so that~$\Alice.x \neq \Bob.x$, and this is reflected in the substitution and renaming operations.
Local variables can be multiply-located, where we write~$\rho.x$ to mean that the local variable~$x$
is in the namespace of every location in the set~$\rho$.

A core feature of choreographies is message passing, which \langname denotes as $C \ChorSend \rho$
to indicate sending the result of evaluating~$C$ to all locations in the set~$\rho$.
Here~$C$ must produce a local value located at~$\ell$ (potentially among others); $\ell$ then sends that value as a message to everyone in $\rho$.
When the result of~$C$ exists only at a single location, we elide the~$\ell$ on the arrow for simplicity.
After the message send, the locations that know the result of~$C$ include~$\rho$ and \emph{anyone who already knew the output of~$C$}.
For example, the choreography $\Alice.3 \ColSend \{\Bob, \Client\}$ reduces to $\{\Alice, \Bob, \Client\}.3$.

Choreographic conditionals, written $\ITE{C}{C_1}{C_2}$, branch the entire choreography on the result of a local computation.
Here~$C$ must produce a boolean known to~$\rho$.
Although the locations in~$\rho$ know which branch to take, other participants in the program may not.
This problem can be solved in two ways:
explicitly share the branch condition with all participants,
or include selection statements~$\syncs{\ell}{d}{\rho'} \seq C$ in the branches to inform locations in~$\rho'$ of which branch was taken.
The former follows the literature on multiply-located values~\citep{SweetDHEHH23,BatesKJSKN25},
while the latter follows literature on selection messages~\citep[see, e.g.,][]{Montesi13,HirschG22,Montesi23,GraversenHM24}.
As the first choreographic language to incorporate both options, \langname allows added flexibility
and supports both of the following equivalent choreographies.
\begin{mathpar}
  \ITE*[\{\Alice,\Bob\}]{\Alice.\True \ColSend \Bob}{\Bob.1}{\Bob.2}
  \and
  \ITE*[\{\Alice\}]{\Alice.\True}{\syncs{\Alice}{\Left}{\Bob} \seq \Bob.1}{\syncs{\Alice}{\Right}{\Bob} \seq \Bob.2}
\end{mathpar}

The type abstraction $\TFun{\alpha \knd \kappa}{C}$ and type application $C~t$ together implement polymorphism.
As described in Section~\ref{sec:kind-system} below, \langname has four kinds,
all of which are valid in type abstractions.

In addition, \langname includes typical algebraic and recursive data types including pairs, projections, injections, $\CaseN$ expressions, and the isorecursive constructor $\FoldN$ and eliminator $\UnfoldN$.
These constructs have the standard semantics for a strict functional language~\citep[Chapters~11 \& 20]{Pierce02}, but note that they represent \emph{global} data.
As a result, after evaluating a choreographic sum type, for instance, \emph{all} participants know and agree on
whether the result is of the form~$\Inl{V_1}$ or~$\Inr{V_2}$, although their knowledge of the contents of~$V_1$ and~$V_2$ may differ.

A major contribution of this work is the presence of \emph{two} \LetN expressions: one for local values and one for types.
Local-let expressions act like standard let expressions, binding local variables to the result of choreographic computations.
If~$C_1$ produces a value located at~$\rho$, then $\LetIn{\rho.x \ty t}{C_1}{C_2}$
binds the result to variable~$x$ in the namespace of~$\rho$, making it available in future local computations.
Importantly, $\rho$~may be any subset of the locations who know the output of~$C_1$.

Our new type-let expressions, written~$\LetIn{\rho.\alpha \knd \kappa}{C_1}{C_2}$,
convert \emph{representations} of locations, location sets, and types, into \emph{actual} locations, location sets, and types.
They are semantically similar to a local-let binding above, except they bind type variables rather than local variables.
Specifically, $C_1$~must produce a representation known to~$\rho$ of a location name, location set, or local type,
which is then reified to a type-level value and bound to~$\alpha$ in the body of~$C_2$.
We elide~$\rho$ and~$\kappa$ for legibility when they are obvious from context.
Combining  the type-let expression with the collecting sends described above allows arbitrary local language computations to dynamically select locations and propagate the choice to the choreographic level,
as shown in Example~\ref{ex:dynamic-locations}.

\begin{ex}[Load Balancer]
  \label{ex:dynamic-locations}
  Recall the example of a distributed thread pool from Section~\ref{sec:introduction}.
  Representations of type $\Loc$ are a value like any other,
  so the local language could provide a function $\Tasks : \Loc \to \Int$ that returns the number of tasks assigned to a worker.
  The (\Mngr)anager can then balance the load between workers~\Alice and~\Bob by allocating computation~$e$ as follows:
  \[
    \LetIn{\alpha}{\Mngr.\left(
    \begin{array}{@{}l@{}}
      \LocalLangFont{if}~(\Tasks(\say{\Alice}) \LocalLess \Tasks(\say{\Bob})) \\
      \LocalLangFont{then}~\say{\Alice}~\LocalLangFont{else}~\say{\Bob}
    \end{array}
    \right) \ColSend \{\Alice,\Bob,\Client\}}{\alpha.e \ColSend \Client}
  \]
  Say~\Alice currently has fewer tasks than~\Bob, so the local program will evaluate to~$\say{\Alice}$,
  meaning the first expression in the let will evaluate to
  \[
    \Mngr.\say{\Alice} \ColSend \{\Alice,\Bob,\Client\} \step \{\Mngr,\Alice,\Bob,\Client\}.\say{\Alice}
  \]
  The type-let then reifies the representation $\say{\Alice}$ to an actual location, producing the step
  \[
    \LetIn{\alpha}{\{\Mngr,\Alice,\Bob,\Client\}.\say{\Alice}}{\alpha.e \ColSend \Client}
    \step
    \Alice.e \ColSend \Client
  \]
\end{ex}

\subsection[Lambda-QC Semantics]{\langname Semantics}
\label{sec:semantics}

The operational semantics of \langname consists of a small-step relation using a labeled transition system of the form~$C_1 \step[R] C_2$.
The label~$R$ represents a redex that tracks the specific reduction occurring.

Choreographies describe the actions of multiple locations,
and so distinct locations should be able to perform (unrelated) actions in any order.
To capture this idea, our semantics includes \emph{out-of-order} reductions.
Importantly, operations for any one location should always execute in the specified order.
To enforce this requirement,
the semantics uses the redices in the step relation to determine which locations are involved in the step,
and only allows reordering of operations when the computation it is jumping before involves a disjoint set of locations.

\begin{figure}
  \begin{syntax}
    \category[Redices]{R}
    \alternative{\RDone{\rho}{e_1}{e_2}}
    \alternative{\RApp}
    \alternative{\RSendV{\ell}{m}{\rho}}
    \alternative{\LetN~\alpha.\rho \ChorDef t}
  \end{syntax}
  \[
    \rulefiguresize
    \def\arraystretch{1.1}
    \begin{array}{l@{\qquad}l}
      \fbox{$\rloc{R}$} & \fbox{$\cloc{C}$} \\
      \begin{array}{r@{\hspace{0.5em}}c@{\hspace{0.5em}}l}
        \rloc{\RDone{\rho}{e_1}{e_2}} & \defeq & \rho \\
        \rloc{\RApp} & \defeq & \Locations \\
        \rloc{\RSendV{\ell}{m}{\rho}} & \defeq & \{\ell\} \cup \rho \\
        \rloc{\RLet{\rho.\alpha}{t}} & \defeq & \rho
      \end{array}
      &
      \begin{array}{r@{\hspace{0.5em}}c@{\hspace{0.5em}}l}
        \cloc{X} & \defeq & \varnothing
        \\
        \cloc{\rho.e} & \defeq & \rho
        \\
        \cloc{\Fun{F}{X}{C}} & \defeq & \varnothing
        \\
        \cloc{C_1~C_2} & \defeq & \Locations
        \\
        \cloc{C \ChorSend[\ell] \rho} & \defeq & \cloc{C} \cup \{\ell\} \cup \rho
        \\
        \cloc{\LetIn{\rho.\alpha \knd \kappa}{C_1}{C_2}} & \defeq & \cloc{C_1} \cup \cloc{C_2} \cup \rho
      \end{array}
    \end{array}
  \]
  \caption{
    Selected Redices and Location Function Rules.
    Here~$m$ is either a local value~$v$ or a selection label~$d$.
  }
  \label{fig:redices}
\end{figure}

More precisely, the \emph{redex locations} function~$\rloc{R}$ returns the set of locations involved in~$R$
and the \emph{choreography locations} function~$\cloc{C}$ gives those involved in~$C$.
For example, if~\Alice sends~$v$ to~\Bob (denoted by the redex $\RSendV{\Alice}{v}{\Bob}$),
then precisely~\Alice and~\Bob participate, so $\rloc{\RSendV{\Alice}{v}{\Bob}} = \{\Alice, \Bob\}$.
The~$\cloc{C}$ function operates on a whole choreography, not just a single step,
so even though~\Alice must take multiple steps before~\Bob gets involved,
$\cloc{\LetIn{\Alice.x}{\Alice.1}{(\Alice.(1 \LocalPlus x) \ColSend \Bob)}} = \{\Alice, \Bob\}$.
Figure~\ref{fig:redices} shows selected redices and definitions of both location functions.

Like in Pirouette, \langname requires global synchronization at function applications---all locations must simultaneously take the same step.
We also require the same for our new choreographic language constructs, including type applications and data eliminators like choreographic case expressions.
To enforce this property, all locations are considered to be involved in a choreography with a function application and the redex associated with applying a function---
$\cloc{C_1~C_2} = \rloc{\RApp} = \Locations$---
and similarly for the other synchronizing syntactic forms.
Note that this synchronization is a technical requirement necessary to prove the choreographic semantics match those of the compiled programs,
and is reflected in the target language as well (see Section~\ref{sec:network-semantics}).

It is then safe to move a step~$R$ before an entire computation~$C$ if the set of participants are disjoint (i.e., $\cloc{C} \cap \rloc{R} = \varnothing$),
even if a normal in-order execution would execute~$C$ before step~$R$.
The following out-of-order rule for type-let expressions exhibits this structure,
also prohibiting the out-of-order step from including locations binding a variable in the let.
\[
  \rulefiguresize
  \CTyLetIRule[left]
\]

\langname also allows out-of-order execution in the branches of an if-expression before fully evaluating the branch condition.
Such a step is safe only when:
(1) the locations involved in the step are disjoint from those computing the branching condition, similarly to \ruleref{C-TyLetI},
and (2) this precise step is guaranteed to happen eventually.
The latter holds only if both branches take identical steps, which we enforce by requiring identical redices.
The result is the following \ruleref{C-IfI} rule.
\[
  \rulefiguresize
  \CIfIRule[left]
\]

As an example, consider the choreography $\ITEBase{\Alice.e}{\Bob.(2 \LocalPlus 3)}{\Bob.(2 \LocalPlus 3)}$.
Although the condition is not yet evaluated, \Bob~will run the same program on either branch, and~\Bob is not involved in computing the branch condition.
That is, no matter what~$\Alice.e$ evaluates to, \Bob~will run~$\Bob.(2 \LocalPlus 3)$.
It is thus safe to reduce both branches to~$\Bob.5$ and the overall choreography to $\ITEBase{\Alice.e}{\Bob.5}{\Bob.5}$.
However, no reordering is possible in the following choreography.
\[
  \ITE[\Alice]{\big(\LetIn{\Alice.x}{(\Bob.5 \ColSend \Alice)}{\Alice.(x \LocalLess 4)}\big)}{\Bob.(2 \LocalPlus 3)}{\Bob.(2 \LocalPlus 3)}
\]
Despite~\Bob performing identical computation in both branches and only~\Alice knowing which way the branch will go,
executing the branches before stepping $\Bob.5 \ColSend \Alice$ in the condition would reorder local operations for~\Bob, which is not permitted.

\begin{figure}
  \begin{ruleset}
    \SetRuleLabelLoc{left}
    \CDoneRule
    \and
    \CAppRule
    \and
     \toggletrue{TyLetVLinebreak}
    \CTyLetVRule
    \and
    \CSendVRule
  \end{ruleset}
  \caption{Selected \langname Operational Semantics}
  \label{fig:semantics}
\end{figure}

Figure~\ref{fig:semantics} contains a selection of additional semantic rules.
\ruleref{C-Done} shows how local programs execute in a choreography,
\ruleref{C-App} applies a function to its argument,
\ruleref{C-TyLetV} shows how type representations are reified to types,
and \ruleref{C-SendV} formalizes the collecting message send semantics discussed in Section~\ref{sec:syntax}.
Each rule in which~$\rho$ appears also requires $\fv{\rho} = \varnothing$,
which simply demands all location variables to be resolved before taking a step.
Otherwise it would be impossible to know who is performing the computation.
Due to the out-of-order rules, this condition is nontrivial, even for closed choreographies.

The remaining rules, which are are either very similar to those presented here or are standard for a strict functional language,
can be found in Appendix~\ref{sec:full-chor-sem}.

\subsubsection{Substitution}
\label{sec:location-substitution}
The separation between choreographic variables, local variables, and type variables
results in three corresponding types of variable substitution.
Choreographic variable substitution, denoted~$\subst{C_1}{X}{C_2}$, follows standard capture-avoiding substitution rules.
Local variable substitution for locations or location sets, denoted~$\hsubst{C}{\ell}{x}{e}$ or~$\hsubst{C}{\rho}{x}{e}$, respectively, is similar,
but it operates only over the namespace of the location~$\ell$ (resp.~$\rho$), which may itself be a type variable.
Notably, a multiply-located variable must be substituted for its entire namespace simultaneously,
so $\hsubst{(\rho.x)}{\rho}{x}{e} = \rho.e$, but $\hsubst{(\{\ell, \ell'\}.x)}{\ell}{x}{e}$ is undefined.

Type substitution, denoted~$\subst{C}{\alpha}{t}$, requires additional care when substituting locations or location sets.
A na\"{i}ve implementation can capture variables due to namespace collisions.
For example, consider the following choreography with a location variable~$\alpha$ and concrete location~\Location.
\[\begin{array}{l}
  C = \begin{array}[t]{@{}l@{}}
    \LetN~\begin{array}[t]{@{}l@{}}
      \alpha.x \ChorDef \alpha.2 \\
      \Location.x \ChorDef \Location.3 \\
      \alpha.y \ChorDef \Location.x \ColSend \alpha \\
    \end{array} \\
    \In~\alpha.(x \LocalPlus y)
  \end{array}
\end{array}\]
Because~$\alpha$ and \Location are (syntactically) distinct, the local variables~$\alpha.x$ and~$\Location.x$ are also distinct,
so this choreography should always produce~$\alpha.5$ regardless of the concrete location to which~$\alpha$ resolves.

But when~$\alpha$ resolves to~\Location, how do we define~$\subst{C}{\alpha}{\Location}$?
One might think that, since nothing binds~$\alpha$  in~$C$, we can simply replace all instances of~$\alpha$ with~\Location.
However, doing so makes the previously-distinct local variables~$\alpha.x$ and~$\Location.x$ collapse.
The variable~$\alpha.x$ is incorrectly captured by the definition of~$\Location.x$ and the choreography would wrongly evaluate to~$\Location.6$.
Note that, when substituting~$\subst{}{\alpha}{\Location}$, capture can occur when binding either~$\alpha.x$ or~$\Location.x$.
Inside a binding of~$\alpha.x$, free instances of~$\Location.x$ in the body will be captured,
and so too will free instances of~$\alpha.x$ be captured when substituting under a binding of~$\Location.x$.

To avoid this namespace capture, substitutions of locations in local-let expressions must rename variables when binding within \emph{either} namespace.
The following rules define
safe location set substitution for local-let bindings, with single location substitution defined as a special case by replacing the set~$\sigma$ with the singleton~$\{\ell\}$
in the side conditions.
Here $\fv[\rho]{C}$ denotes the local variables free in~$C$ for \emph{any} location in~$\rho$.
\[
  \addtocounter{numlevels}{1}
  \rulefiguresize
  \subst{(\LetIn{\rho.x}{C_1}{C_2})}{\alpha}{\sigma} = \begin{cases}
    \LetIn*{(\subst{\rho}{\alpha}{\sigma}).x}{\subst{C_1}{\alpha}{\sigma}}{\subst{C_2}{\alpha}{\sigma}}
    & \text{if}~\alpha \subseteq \rho ~\text{and}~ x \notin \fv[\sigma - \rho]{C_2}
    \\[2em]
    \LetIn*{(\subst{\rho}{\alpha}{\sigma}).y}{\subst{C_1}{\alpha}{\sigma}}{\subst{\hsubst{C_2}{\rho}{x}{y}}{\alpha}{\sigma}}
    & \begin{array}[t]{@{}l@{}}
      \text{if}~\alpha \subseteq \rho, x \in \fv[\sigma - \rho]{C_2}, \\
      \text{and}~y \notin \fv[\rho \cup \sigma]{C_2}
    \end{array}
    \\[2em]
    \LetIn*{\rho.y}{\subst{C_1}{\alpha}{\sigma}}{\subst{\hsubst{C_2}{\rho}{x}{y}}{\alpha}{\sigma}}
    & \begin{array}[t]{@{}l@{}}
      \text{if}~\alpha \nsubseteq \rho, \sigma \cap \rho \neq \varnothing, \\
      x \in \fv[\alpha]{C_2},
      \text{and}~y \notin \fv[\rho \cup \alpha]{C_2}
    \end{array}
    \\[2em]
    \LetIn*{\rho.x}{\subst{C_1}{\alpha}{\sigma}}{\subst{C_2}{\alpha}{\sigma}}
    & \begin{array}[t]{@{}l@{}}
      \text{if}~\alpha \nsubseteq \rho ~\text{and} ~\text{either} \\
      \sigma \cap \rho = \varnothing ~\text{or}~ x \notin \fv[\alpha]{C_2}
    \end{array}
  \end{cases}
  \addtocounter{numlevels}{-1}
\]


\subsection[Lambda-QC Kinding System]{\langname Kinding System}
\label{sec:kind-system}

We now turn to our static semantics, which has two components: an elementary kinding system, and a typing system.
For notational brevity we assume throughout that new variables are always fresh.
A kinding judgment takes the form~$\chorkinded{\Gamma}{t}{\kappa}$, where~$\Gamma$ is a kinding context,~$t$ is a type, and~$\kappa$ is a kind.
The kind~$\kappa$ classifies~$t$ as either a program type~($*$), a location~($*_\ell$), a set of locations~($*_s$), or a local program type~($*_e$).
Figure~\ref{fig:types} presents the syntax for the \langname types and kinds.

\begin{figure}[b]
  \begin{syntax}
    \category[Kinds]{\kappa}
    \alternative{*}
    \alternative{*_\ell}
    \alternative{*_s}
    \alternative{*_e}


    \category[Local Program Types]{t_e}
    \alternative{\alpha}
    \alternative{\Bool}
    \alternative{\Typ}
    \alternative{\Loc_\rho}
    \alternative{\LocSet_\rho}
    \alternative{\ldots}

    \categoryFromSet[Locations]{\Location,\Alice,\Bob,\ldots}{\Locations}

    \category[Choreography Types]{\tau, \ell, \rho, t}
    \alternative{\alpha}
    \alternative{t_e @ \rho}
    \alternative{\tau_1 \rightarrow \tau_2}
    \alternative{\forall \alpha \knd \kappa.\tau}
    \\
    \alternative{\tau_1 \times \tau_2}
    \alternative{\tau_1 + \tau_2}
    \alternative{\mu \alpha.\tau}
    \alternative{\Location}
    \alternative{\{\ell\}}
    \alternative{\rho_1 \cup \rho_2}
  \end{syntax}

  \caption{Syntax of Types and Kinds. Here~$\alpha$ is a type variable.}
  \label{fig:types}
\end{figure}

All program types of kind~$*$ directly extend their analogues from System~F, except for base types.
Our base types take the form~$t_e @ \rho$ and represent a single local program of type~$t_e$ running at all locations in the set~$\rho$.
The kind~$*_\ell$ represents location names, which can refer to either concrete locations~$\Location \in \Locations$, or in-context location variables.
The kind~$*_s$ classifies (non-empty) finite sets of location names, which can be either a type variable, a singleton set~($\{\ell\}$), or a union of sets~($\rho_1 \cup \rho_2$).
Finally, types of kind~$*_e$ are precisely the types included in the local language under a given type variable context.
Notably, local types may use type variables bound in the choreography, even if the local type system does not include a type variable binding mechanism.

The full kinding rules can be found in Appendix~\ref{sec:full_kinds}.

\subsection[Lambda-QC Type System]{\langname Type System}
\label{sec:type-system}
The second component of our static semantics is the type system.
Typing judgments for \langname take the form~$\chortyped{\Gamma}{\Delta}{\Sigma}{C}{\tau}$,
where~$\Gamma$ is a kinding context, as described above,
$\Delta$~is a choreographic typing context that handles variables bound by choreographic functions and case expressions,
and~$\Sigma$ is a local typing context.
These contexts are handled in a standard manner for a polymorphic type system~\citep[Chapter~23]{Pierce02}.
The local typing context~$\Sigma$ is a list of ascriptions of the form~$\rho.x \ty t_e$
tracking variables bound by local let expressions (see Section~\ref{sec:syntax}).
Figure~\ref{fig:selected-type-rules} contains a selection of typing rules for \langname.

\begin{figure}
\begin{ruleset}
  \SetRuleLabelLoc{left}
  \TDoneRule
  \and
  \TSendRule
  \and
  \TLetLocalRule
  \and
  \TLetLocRule
\end{ruleset}
  \caption{Selected Typing Rules}
  \label{fig:selected-type-rules}
\end{figure}

The \ruleref{T-Done} rule type-checks local computations by appealing to the local type system, but it needs to know which local variables are in scope.
In a multiply-located computation~$\rho.e$, every location in~$\rho$ should compute the same result,
so each free variable in~$e$ must be bound with the same meaning at every location in~$\rho$.
To check this requirement, we define a projection operation $\proj{\Sigma}{\rho}$ that restricts~$\Sigma$
to only those local variables bound in a namespace $\rho'$ where $\rho \subseteq \rho'$.
Formally,
\[
  \proj{\Sigma}{\rho} \defeq
  \begin{cases}
    \cdot &\text{if}~\Sigma = \cdot\\
    \proj{\Sigma'}{\rho}, x \ty t_e &\text{if}~\Sigma = \Sigma', \rho'.x \ty t_e ~\text{and}~\rho \subseteq \rho'\\
    \proj{\Sigma'}{\rho} &\text{if}~\Sigma = \Sigma', \rho'.x \ty t_e ~\text{and}~\rho \not\subseteq \rho'
  \end{cases}
\]

The \ruleref{T-Send} rule ensures (1) that only local values can be sent by requiring~$C$ to have type~$t_e @ \rho_1$,
and (2) that the sender~($\ell$) must know the value being sent.
By locating the output type at $\rho_1 \cup \rho_2$, it captures the multiply-located collecting semantics described in Section~\ref{sec:semantics}.

The other two rules in Figure~\ref{fig:selected-type-rules} are the two forms of let binding.
The \ruleref{T-LetLocal} rule for local-let bindings is very similar to a traditional let binding rule.
It just restricts to binding local values---$C_1$ has type $t_e @ \rho_2$---and requires that
$\rho_1$ (the locations binding~$x$) must all have access to the result of the computation~$C_1$---that is, $\rho_1 \subseteq \rho_2$.

The \ruleref{T-LetLoc} rule for type-let expressions, core to supporting first-class location names, is much more subtle.
Indeed, a na\"{i}ve definition could introduce type dependency or unsoundness.
Type dependency is the simpler concern, and is addressed by the premise $\chorkinded{\Gamma}{\tau}{*}$,
which demands that~$\tau$ be a well-formed type in the context~$\Gamma$ \emph{without the newly-bound variable~$\alpha$}.
As a result, while the body~$C_2$ of the let may freely reference~$\alpha$, its \emph{type}~$\tau$ may not.
This restriction prevents the type of the entire let expression from depending on the value of~$C_1$, thereby avoiding value dependency in the type system.
As an example, the program below on the left is not well-typed, because the body has type~$\Int @ \alpha$ in which~$\alpha$ is free.
By contrast, the program on the right is well-typed because the body, and the choreography itself, has the type~$\Int @ \Bob$.
\begin{mathpar}
  {}\nproves
  \left(\LetIn*{\Alice.\alpha}{\Alice.\say{\Alice}}{\alpha.(1 \LocalPlus 1)}\right)
  : \tau
  \and
  {}\proves
  \left(~
    \begin{array}[c]{@{}l@{}}
      \LetN\begin{array}[t]{l@{}l@{}}
        \{\Alice,\Bob\}.\alpha~&\ChorDef \{\Alice,\Bob\}.\say{\Alice} \\
        \Bob.x~&\ChorDef \alpha.(1 \LocalPlus 1) \ColSend \Bob
      \end{array} \\
      \In~\Bob.x
    \end{array}
  ~\right)
  : \Int @ \Bob
\end{mathpar}

The other unusual premise to \ruleref{T-LetLoc} is~$\rho_1 \subseteq \rho_2 \subseteq \rho_3$, which is necessary to ensure soundness.
To understand why, recall Example~\ref{ex:dynamic-locations}, in which the thread-pool manager~\Mngr dynamically assigns a task to either~\Alice or~\Bob.
Consider what happens if~\Mngr were to inform only~\Bob and the client~\Client of their decision, but not~\Alice, the selected worker.
As~\Alice is not aware that she should compute~$e$ and send back the result,~\Client will wait forever on a message from~\Alice that never arrives, causing a deadlock.

Amazingly, the simple premise~$\rho_1 \subseteq \rho_2 \subseteq \rho_3$ prevents all deadlocks resulting from the type-let expression.
As described in Section \ref{sec:local-types}, $\rho_1$~is a static upper bound on the set of values to which the dynamically generated location may resolve.
The location variable is bound in the namespace of~$\rho_2$, so only those locations may use the result,
and all locations in~$\rho_3$ know the value of the dynamically generated location name.
The subset relationship therefore ensures that any location to which~$\alpha$ may resolve will bind~$\alpha$,
and that all locations that bind~$\alpha$ know the value.
In the unsound hypothetical above, $\rho_1 = \{\Alice,\Bob\}$ and $\rho_2 = \{\Mngr,\Bob,\Client\}$, violating this premise.

The full type system---available in Appendix~\ref{sec:full_chor_types}---%
is sound with respect to the operational semantics, demonstrated by the following mechanically verified theorems.

\begin{thm}[Type Preservation]\label{thm:preservation}
  If $\chortyped{\Gamma}{\Delta}{\Sigma}{C_1}{\tau}$ and $C_1 \step C_2$, then $\chortyped{\Gamma}{\Delta}{\Sigma}{C_2}{\tau}$.
\end{thm}
\begin{thm}[Progress]\label{thm:progress}
  If $\choremptyped{C_1}{\tau}$, then either~$C_1$ is a value, or there is some~$C_2$ such that $C_1 \step C_2$.
\end{thm}


\subsection{Example: Bookstore Aggregator}
\label{ex:bookstore-aggregator}
To see how the added features of \langname support more interesting systems,
we present a generalization of the classic bookseller protocol
that models a storefront with arbitrarily many sellers.
In this example, the system maintains an associative list---the \Inventory argument, encoded as a choreographic recursive type---of books for sale, their price,
and who is selling them, with data stored at an aggregator~\Mngr.
To buy a book, a client~$C$ sends~\Mngr the title, who then looks it up in the \Inventory
using a choreographic lookup function
\[ \Lookup : \forall L, \tau \ldotp \String @ L \to \List{\String @ L \times \tau} \to \Maybe{\tau} \]
whose definition is standard and omitted here for brevity.
If the book is not found, the process ends.
If the book is found, \Mngr~notifies~$C$ and the various \Shops of the identity of the seller, which is bound to the variable~$S$.
The aggregator~\Mngr then sends the price to the buyer and the seller, the buyer decides if the price is acceptable, and if so, they complete the transaction
by calling the function
\[ \Transact : \forall C, S \ldotp \String @ \{C,S\} \to \Int @ \{C,S\} \to (), \]
which we assume is a common interface provided by the storefront.
\begin{align*}
  & \Aggregator : \forall C \ldotp \String @ C \to \Int @ C \to \List{\String @ \Mngr \times \Int @ \Mngr \times \Loc_\Shops @ \Mngr} \to ()\\
  & \Aggregator~C~\Title~\Budget~\Inventory = \\
  & \qquad
  \def\arraystretch{1.1}
  \RawCase*{(\Lookup~\Mngr~(\Int @ \Mngr \times \Loc_\Shops @ \Mngr)~(\Title \ColSend \Mngr)~\Inventory)}{\None}{()}
    {\Some{(\Mngr.\Price, \Mngr.\SellerName)}}{
        \MultiLet{{S}{\Mngr.\SellerName \ColSend \{C\} \cup \Shops}
                  {\{\Mngr,C,S\}.\Price}{\Mngr.\Price \ColSend \{C,S\}}
                  {\{C,S\}.\Buy}{C.(\Price \LocalLessEq \Budget) \ColSend \Seller}}
                 {\ITEBase*{\{C,S\}.\Buy}{\Transact~C~S~(\Title \ColSend S)~\{C,S\}.\Price}{()}}
    }
\end{align*}

While we do not detail the full proof here, it is straightforward to check that the above code is well-typed using the type system presented above.


\section{Network Language}
\label{sec:network-lang}

Choreographies specify the behavior of concurrent systems,
and compiling them to a set of programs that can actually run concurrently at different locations requires a language to specify that system.
We thus provide a \emph{network language} for \langname.
It specifies programs at individual locations and how to compose them into a parallel system with concurrent operational semantics.

\subsection{Network Language Syntax}
\label{sec:network-syntax}

\begin{figure}[b]
  \begin{syntax}
    \category[Network Program]{E}
    \alternative{X} \alternative{\Ret{e}} \alternative{\CtrlUnit} \alternative{\SendTo{E}{\rho}} \alternative{\RecvFrom{\ell}} \\
    \alternative{E_1 \CtrlSeq E_2}
    \alternative{\CtrlFun{F}{X}{E}} \alternative{E_1~E_2} \alternative{\CtrlTFun{\alpha}{E}} \alternative{E~t} \\
    \alternative{\CtrlLetIn{x}{E_1}{E_2}} \alternative{\CtrlLetIn{\alpha \knd \kappa}{E_1}{E_2}} \\
    \alternative{\AllowChoice{\ell}{E_1}{E_2}}
    \alternative{\ChooseFor{d}{\rho}{E}} \\
    \alternative{\CtrlITE{E}{E_1}{E_2}}
    \alternative{\AmIIn{\rho}{E_1}{E_2}}
    \category[Systems]{\Pi}
    \alternative{L_1 \triangleright E_1 \mathrel{\parallel} \ldots \mathrel{\parallel} L_n \triangleright E_n}\\
  \end{syntax}

  \caption{Selected Network Program Syntax. Here $L \in \Locations$ is a concrete location name.}
  \label{fig:network-lang-syntax}
\end{figure}

The network language is a concurrent $\lambda$-calculus where messages are values from the same local language as choreographies.
The syntax is similar to the choreographic syntax from Section~\ref{sec:language}, except sending and receiving messages are now split into two separate constructs.
Figure~\ref{fig:network-lang-syntax} contains selected syntactic forms, using the same variable naming conventions as choreographies.
The full definition is available in Appendix~\ref{sec:full-ntwk-syntax}.

The network language counterpart of the choreographic local computation~$\rho.e$
is the return expression~$\Ret{e}$, which similarly executes the local language program~$e$.
Since a network program is only for a single location~$L$, it should only run~$e$ when $L \in \rho$ and do nothing otherwise.
We include a unit value~$\CtrlUnit$ to represent a network program that does nothing.

Choreographies have a single message sending operation, which represents both sending and receiving the message.
As the network language describes only a single location's behavior, it splits these into two constructs:
$\SendTo{E}{\rho}$ which multicasts the results of~$E$ to every location in~$\rho$,
and $\RecvFrom{\ell}$, which receives a local value from location~$\ell$.
Note that \SendN can only send local values, so $\SendTo{\CtrlUnit}{\rho}$, for instance, would be stuck.

Both local-let and type-let expressions are mirrored identically from choreographies to the network language,
as are recursive functions, type abstractions, and applications for both.
The network language also includes a primitive sequencing operator $E_1 \CtrlSeq E_2$
for a similar reason to why it includes~$\CtrlUnit$.
Specifically, a location~$L$ may have computation to perform in the head and body of a choreographic let expression,
yet not be included in the locations binding a variable.
In this case, $L$~will sequence their actions from the two parts of the let expression.

Finally, the network language contains three branching mechanisms for different purposes:
the standard $\CtrlITE{E}{E_1}{E_2}$ construct,
\AllowChoiceN to implement choreographic selection statements,
and $\AmIInN$, which branches based on the identity of the current location.

Recall from Section~\ref{sec:syntax} that a selection statement $\syncs{\ell}{d}{\rho} \seq C$
has~$\ell$ send~$d$ to each location in~$\rho$ so they know which branch to take in a choreographic \IfN statement.
As with message sends, the network language splits the construct in two: sending and receiving.
Each recipient is waiting on an \emph{external choice}.
Some other location~$\ell$ must pick either~$\Left$ or~$\Right$, represented by the expression $\AllowChoice{\ell}{E_1}{E_2}$,
which will execute~$E_1$ after receiving~$\Left$ or~$E_2$ after receiving~$\Right$.
Meanwhile, $\ell$~knows the choice~$d$ and can send it to all locations in~$\rho$ before executing~$E$ using the expression $\ChooseFor{d}{\rho}{E}$.

Note that an \AllowChoiceN with only one branch, such as $\AllowOneChoice{\ell}{\Left}{E}$, is valid.
This one-sided branch will execute~$E$ if~$\ell$ sends~$\Left$, but will get stuck if~$\ell$ sends~$\Right$.
We write~$E_\bot$ for the program in a branch that might be missing.

The last form of branching is the ``AmI-In'' expression $\AmIIn{\rho}{E_1}{E_2}$, which conditions on whether the currently executing location is in~$\rho$.
That is, when executing at location~$L$, $E_1$~will execute if~$L \in \rho$, and~$E_2$ will execute otherwise.
This construct generalizes the ``AmI'' expression of PolyChor$\lambda$~\citep{GraversenHM24},
which uses a single location~$\ell$ instead of a set~$\rho$, and branches based on equality with~$\ell$ rather than inclusion.
We write~$\AmI{\ell}{E_1}{E_2}$ as a shorthand for~$\AmIIn{\{\ell\}}{E_1}{E_2}$.

\subsection{Network Language Operational Semantics}
\label{sec:network-semantics}

\begin{figure}
  \begin{syntax}
    \category[Transition Labels]{l}
    \alternative{\iota} \alternative{\iotasync}
    \alternative{\RSendNtwk{m}{\rho}} \alternative{\RRecvNtwk{L}{m}}
  \end{syntax}

  \begin{ruleset}
    \SetRuleLabelLoc{left}
    \NRetRule
    \and
    \NAppRule
    \and
    \NSendRule
    \and
    \NRecvRule
    \and
    \NChooseRule
    \and
    \toggletrue{NAllowLLinebreak}
    \NAllowLRule
  \end{ruleset}
  \caption{Selected Network Language Operational Semantics}
  \label{fig:ntwk-semantics}
\end{figure}

The network language semantics is a labeled transition system $L \triangleright E_1 \ctrlstep{\smash{\scriptstyle l}} E_2$
where~$L$ is the location executing the program and~$l$ is the label on the step.
Just as network programs specify the operations of a single location, the labels only acknowledge the view of that location.
Figure~\ref{fig:ntwk-semantics} shows selected labels and rules,
with the full semantics available in Appendix~\ref{sec:full-ntwk-sem}.

The $\iota$~label denotes an internal step,
where the reduction does not require interaction between locations.
The \ruleref{N-Ret} rule for local steps shows an example use of~$\iota$.

The \emph{synchronized} internal label~$\iotasync$ denotes internal steps where \emph{all} locations must take that step at the same time.
It is used for steps like $\beta$-reduction (\ruleref{N-App}) where
the corresponding choreographic step modifies the choreography for all participants.
Synchronization in these cases is a technical requirement needed to prove correctness of the system,
which we compare to other approaches in Section~\ref{sec:choreo-prog}.

The labels~$\RSendNtwk{m}{\rho}$ and~$\RRecvNtwk{L}{m}$ appear with message sends and receives, respectively, including selection messages.
As recipients do not know the value in advance, \ruleref{N-Recv} is non-deterministic, allowing any value to arrive.
The system semantics below ensures senders and receivers agree on the message.
The same is true for \ruleref{N-AllowL} and a symmetric \textsc{N-AllowR} rule.
The sender's perspective is governed by \ruleref{N-Send} and \ruleref{N-Choose}, which ensure that the transition label matches the message and recipients intended by the program.

\subsubsection{Network Systems}
\label{sec:system-semantics}

\begin{figure}
  \begin{syntax}
    \category[System Label]{l_S}
    \alternative{\iota} \alternative{\iotasync} \alternative{L_1.m \sendsto \rho}
  \end{syntax}

  \begin{ruleset}
    \Rule{Internal}{
      L \triangleright \Pi(L) \ctrlstep{\iota} E}
    {\Pi \systemstep[\iota] \subst{\Pi}{L}{E}}
    \and
    \Rule{Sync-Internal}{
      \forall L \in \text{dom}(\Pi)\ldotp \Big(L \triangleright \Pi(L) \ctrlstep{\iotasync} \Pi'(L)\Big)}
    {\Pi \systemstep[\iotasync] \Pi'}
    \and
    \Rule{Comm}{
      L_1 \notin \rho \\
      L_1 \triangleright \Pi(L_1) \ctrlstep{\RSendNtwk{m}{\rho}} E_1 \\\\
      \forall L \in \rho\ldotp \Big(L \triangleright \Pi(L) \ctrlstep{\RRecvNtwk{L_1}{m}} E_L\Big) \\
    }{\Pi \systemstep[L_1.m \sendsto \rho] \subst*{\Pi}{{L_1}{E_1}{\rho}{E_L}}}
  \end{ruleset}
  \caption{System Semantics and Labels}
  \label{fig:system-semantics}
\end{figure}

A choreography specifies \emph{interactions} between multiple locations,
which we represent with a parallel composition of network programs at different locations.
Formally, a \emph{system} $\Pi ={} {\parallel_{L \in \SysLocs} (L \triangleright E_L)}$
maps each location~$L$ in a finite set $\SysLocs \subseteq \Locations$ to the network program~$E_L$ it is currently executing.
The system semantics is itself a larger labeled transition system, with combined labels shown in Figure~\ref{fig:system-semantics}.
While network program labels show only one side of a send or receive, system labels reflect both sides
and the semantics ensures that senders and recipients agree on the content of a message.

Figure~\ref{fig:system-semantics} also shows how to lift the operational semantics of the network language to systems with three rules.
\ruleref{Internal} allows one location to independently take an internal step.
\ruleref{Sync-Internal} allows all locations to simultaneously perform a synchronized step.
\ruleref{Comm} parties sending and receiving messages together,
requiring the sender and \emph{all} specified recipients step at the same time with the same message value.
Notationally, $\subst{\Pi}{\rho}{E_L}$ denotes the updated system mapping~$L$ to~$E_L$ if $L \in \rho$ and $\Pi(L)$ otherwise.


\section{Endpoint Projection}\label{sec:endpoint-projection}
With the compilation target fixed, we turn to the endpoint projection procedure
that defines how to compile a choreography into a concurrent network system.

\subsection{Network Program Merging}\label{sec:network-merge}

It is important to keep the projection definition \emph{compositional} for simplicity and scalability,
but choreographic \IfN statements complicate this process and necessitate a \emph{merge operator}.
To understand why, consider the following choreography~$C$.
\begin{center}
  \begin{tikzpicture}[node distance=0pt]
    \node (If) at (0,0) {$\IfN~\Alice.e~\ThenN$};
    \node[base right=of If] (C1) {$(\syncs{\Alice}{\Left}{\Bob} \seq \Bob.1)$};
    \node[base right=of C1] (Else) {$\ElseN$};
    \node[base right=of Else] (C2) {$(\syncs{\Alice}{\Right}{\Bob} \seq \Bob.2)$};

    \node[below=2.25em of C1] (C1Proj) {$\AllowOneChoice*{\Alice}{\Left}{\Ret{1}}$};
    \node[below=2.25em of C2] (C2Proj) {$\AllowOneChoice*{\Alice}{\Right}{\Ret{2}}$};

    \node[right=of C1Proj, left=of C2Proj] (Merge) {$\Merge$};
    \node[right=of C2Proj] (Eq1) {$=$};
    \node[right=0.75em of Eq1.base] (MergeProj) {$\AllowChoice*{\Alice}{\Ret{1}}{\Ret{2}}$};
    \node (Eq2) at (Eq1 |- C2) {$=$};
    \node (C) at (MergeProj |- Eq2) {$C$};

    \draw[mapsto] (C1) -- (C1Proj) node[label,midway,left,yshift=1.5pt]{$\epp{\cdot}{\Bob}$};
    \draw[mapsto] (C2) -- (C2Proj) node[label,midway,left,yshift=1.5pt]{$\epp{\cdot}{\Bob}$};
    \draw[mapsto] (C) -- (MergeProj) node[label,midway,left,yshift=1.5pt]{$\epp{\cdot}{\Bob}$};
  \end{tikzpicture}
\end{center}
In the \ThenN branch, \Alice~will always send~\Bob the selection message~\Left.
In the projection of this branch, \Bob~should therefore wait to receive~\Left and then return~$1$.
There are no instructions for what to do if~\Bob receives~\Right in this branch---which will never happen---so that side of the \AllowChoiceN is empty.
Similarly, in the \ElseN branch, \Bob~can only ever receive~\Right, so the~\Left side of the \AllowChoiceN is empty.

\begin{figure}
  \begin{ruleset}
    \left(\AllowOneChoice*{\ell}{\CtrlLeft}{E_1}\right) \Merge \left(\AllowOneChoice*{\ell}{\CtrlRight}{E_2}\right) \defeq \AllowChoice*{\ell}{E_1}{E_2}
    \and
    \left(\AllowOneChoice*{\ell}{\CtrlLeft}{E_1}\right) \Merge \left(\AllowOneChoice*{\ell}{\CtrlLeft}{E_1'}\right) \defeq \AllowOneChoice*{\ell}{\CtrlLeft}{E_1 \Merge E_1'}
    \and
    \left(\AllowOneChoice*{\ell}{\CtrlLeft}{E_1}\right) \Merge \left(\AllowChoice*{\ell}{E_1'}{E_2}\right) \defeq \AllowChoice*{\ell}{E_1 \Merge E_1'}{E_2}
  \end{ruleset}
  \caption{Selected Merge Operator Definitions}
  \label{fig:merge-defs}
\end{figure}

In the full \IfN statement, however, both branches are possible, so \Bob's~projection must include both options.
To compositionally combine the \AllowChoiceN statements from each branch, we use a \emph{merge} operator $E_1 \Merge E_2$.
Specifically, $\Merge$ is an idempotent partial binary function defined structurally homomorphically on matching network programs
that incorporates \AllowChoiceN branches that exist on only one side (and homomorphically merges those that exist in both).
Figure~\ref{fig:merge-defs} shows the most interesting rules defining merge, with the full definition available in Appendix~\ref{sec:ntwk-merge-def}.

This merge operator is extremely similar to the one in Pirouette~\citep{HirschG22},
but we allow functions to merge if their bodies merge, not just if their bodies are syntactically equal.

\subsection{Endpoint Projection Definition}\label{sec:epp-defined}
Using this merge function, we can define endpoint projection~(EPP).
The projection of a choreography~$C$ to a location (endpoint)~$L$, denoted $\epp{C}{L}$, is the network program~$L$ runs to implement its part of~$C$.
EPP is partial (i.e., a choreography may fail to project) both because the merge operator is partial and because it must verify that locations only use type variables which they have bound.

\begin{figure}
  \rulefiguresize
  \begin{align*}
    \epp{\rho.e}{L} & \defeq
      \begin{cases}
        \Ret{e} & \text{if}~L \in \rho\\
        \CtrlUnit & \text{otherwise}
      \end{cases}
    \\[\vertrulegap]
    \epp{C \ChorSend \rho}{L} & \defeq
    \begin{cases}
      \SendTo{\epp{C}{L}}{\rho} & \text{if}~L = \ell\\
      \epp{C}{L} \seqfun \RecvFrom{\ell} & \text{if}~L \neq \ell ~\text{and}~ L \in \rho\\
      \epp{C}{L} & \text{otherwise}
    \end{cases}
    \\[\vertrulegap]
    \epp{\syncs{\ell}{d}{\rho} \seq C}{L} & \defeq
    \begin{cases}
      \ChooseFor{d}{\rho}{\epp{C}{L}} & \text{if}~L = \ell \\
      \AllowOneChoice{\ell}{d}{\epp{C}{L}} & \text{if}~L \neq \ell ~\text{and}~ L \in \rho \\
      \epp{C}{L} & \text{otherwise}
    \end{cases}
    \\[\vertrulegap]
    \epp{\ITE[\rho]{C}{C_1}{C_2}}{L} & \defeq
    \begin{cases}
      \CtrlITE{\epp{C}{L}}{\epp{C_1}{L}}{\epp{C_2}{L}} & \text{if}~L \in \rho\\
      \epp{C}{L} \seqfun (\epp{C_1}{L} \Merge \epp{C_2}{L}) & \text{otherwise}
    \end{cases}
    \\[\vertrulegap]
    \epp{\TFun{\alpha \knd \kappa}{C}}{L} & \defeq \begin{cases}
      \CtrlTFun{\alpha \knd *_\ell}{\AmI{\alpha}{\epp{\subst{C}{\alpha}{L}}{L}}{\epp{C}{L}}} & \text{if}~ \kappa = *_\ell \\
      \CtrlTFun{\alpha \knd *_s}{\AmIIn{\alpha}{\epp{\subst{C}{\alpha}{\{L\} \cup \alpha}}{L}}{\epp{C}{L}}} & \text{if}~ \kappa = *_s \\
      \CtrlTFun{\alpha \knd \kappa}{\epp{C}{L}} & \text{otherwise}
    \end{cases}
    \\[\vertrulegap]
    \addtocounter{numlevels}{1}
    \epp{\LetIn{\rho.\alpha \knd *_\ell}{C_1}{C_2}}{L} & \defeq
    \begin{cases}
      \CtrlLetIn*{\alpha \knd *_\ell}{\epp{C_1}{L}}{\AmI{\alpha}{\epp{\subst{C_2}{\alpha}{L}}{L}}{\epp{C_2}{L}}} & \text{if}~L \in \rho\\
      \epp{C_1}{L} \seqfun \epp{C_2}{L} & \text{if}~L \notin \rho ~\text{and}~ \alpha \notin \fv{\epp{C_2}{L}}\\
      \Undef & \text{otherwise}
    \end{cases}
    \addtocounter{numlevels}{-1}
  \end{align*}
  \caption{Selected EPP Definitions}
  \label{fig:epp-defs}
\end{figure}

EPP is defined as a structurally recursive function over the syntax of the choreography.
Most of the rules simply convert choreographic syntax into its network language counterpart,
but in some cases more involved translation is required.
Figure~\ref{fig:epp-defs} shows these cases.

The first four projection definitions capture intuitions described previously.
For $\rho.e$, only locations in~$\rho$ should compute~$e$ while others should do nothing.
For $C \ChorSend \rho$, everyone should perform their portion of the computation specified by~$C$,
then location~$\ell$ should send a message, locations in~$\rho$ should receive a message, and that is all.
For selection messages, location~$\ell$ should announce the choice, while other locations in~$\rho$ wait for it
and condition their behavior on the result.
As there is only one choice here, the \AllowChoiceN only has one branch.
The other branch can appear when projecting conditionals,
where everyone first performs any computation specified by the condition,
then locations who know the result of that condition---locations in~$\rho$---simply branch,
while EPP combines the branches for other locations using the merge operation defined above.
Merging can fail, so EPP also fails for this location if it is undefined.

\para{Sequencing Function}
Note that the rules do not directly use the network language's native sequencing primitive~$E_1 \CtrlSeq E_2$,
but instead use a \emph{collapsing sequencing function}~$E_1 \seqfun E_2$ defined by
\[
  E_1 \seqfun E_2 = \begin{cases}
    E_2 & \text{if}~\CtrlVal{E_1}\\
    E_1 \CtrlSeq E_2 & \text{otherwise}.
  \end{cases}
\]
This can be seen as a peephole optimization to eliminate null sequences
whose primary purpose is to enable projected systems to properly simulate out-of-order choreographic steps.
For instance, without collapsing sequenced values, \Bob's projection of the choreography
$\LetIn{\Alice.x}{\Alice.e}{\Bob.(2 \LocalPlus 3)}$ would be $\CtrlUnit \CtrlSeq \Ret{2 \LocalPlus 3}$.
An out-of-order step at~\Bob would reduce to $\LetIn{\Alice.x}{\Alice.e}{\Bob.5}$,
but to simulate this step, the projection would need to reduce to $\CtrlUnit \CtrlSeq \Ret{5}$,
which is impossible because the reduction occurs to the right of a semicolon.
The actual EPP defined above, however, projects these choreographies to~$\Ret{2 \LocalPlus 3}$ and $\Ret{5}$, respectively, eliminating the concern.

Collapsing values also adds flexibility to EPP.
To see why, consider the following choreography.
\[C = \ITEBase{\Alice.e}{(\LetIn{\Alice.x}{\Alice.(1 + 1)}{\Bob.3})}{\Bob.3}\]
Intuitively, \Bob~takes identical actions---returning $3$---in both branches, so $\epp{C}{\Bob}$~should simply be~$\Ret{3}$.
However, without the sequencing function, the \ThenN branch would project to~$\CtrlUnit \CtrlSeq \Ret{3}$, while the \ElseN branch would project to~$\Ret{3}$.
These network programs do not merge, so $C$~would not project.
Collapsing the \ThenN branch to~$\Ret{3}$ allows the expected projection.

\para{Process Polymorphism}
The final two projections in Figure~\ref{fig:epp-defs} concern location (set) abstraction.
Choreographic type abstractions project to network type abstractions.
When the type variable is a location~($*_\ell$) or location set~($*_s$), however,
the body must behave differently depending on how the location executing relates to the value the type variable resolves to.
\citet{GraversenHM24} solve this problem with the~\AmIN construct,
executing~$\epp{\subst{C}{\alpha}{L}}{L}$ when the locations match and~$\epp{C}{L}$ when they do not.
When the locations match, this projection is clearly correct.
When they do not, the projection is correct because EPP treats location variables as abstract identifiers that are equal only to themselves,
meaning $\epp{C}{L}$ will behave as though $\alpha \neq L$.

Location \emph{set} abstraction generalizes this idea.
Replacing \AmIN with $\AmIInN$ is straightforward, as is $\epp{C}{L}$ treating~$\alpha$ as an abstract set where $L \notin \alpha$.
The other branch, where~$\alpha$ resolves to a set containing~$L$, is more subtle.
This projection of~$C$ must treat~$\alpha$ as though it contains~$L$, \emph{without changing the meaning of~$\alpha$}.
We accomplish this by explicitly adding~$L$ to the set in this branch and using $\epp{\subst{C}{\alpha}{\{L\} \cup \alpha}}{L}$.
The symbolic containment relation recognizes that $L \in \{L\} \cup \rho$ for any~$\rho$, including the variable~$\alpha$,
so the projection will properly treat~$L$ as being in the set.
Moreover, despite symbolically changing the set, the operational semantics remain correct.
This branch only executes when~$\alpha$ resolves to some~$\rho$ where $L \in \rho$, in which case $\{L\} \cup \rho = \rho$.

Finally, type-let expressions combine a structurally recursive projection, also used in local-let expressions, with the location-based conditionals of type abstraction.
For locations binding the type variable, the projection first computes the type, binds it, and then uses \AmIN to condition based on its value.
For other locations, EPP simply sequences the head of the \LetN with its body.
Notably, the type system, to avoid dependency, does not prevent locations outside of~$\rho$ from referencing~$\alpha$.
Any such invalid reference would leave~$\alpha$ unbound in the body of the let, so EPP checks that $\alpha \notin \fv{\epp{C_2}{L}}$.
If both~$\alpha$ is free and $L \notin \rho$, there is no correct way for~$L$ to execute its part of~$C_2$, so the projection is undefined.
The projection for type-let expressions where~$\alpha$ has kind~$*_s$ are nearly identical,
with $\AmIInN$ in place of~\AmIN and the approach described above for type abstractions when $L \in \rho$.

This EPP definition gives the network program corresponding to one location.
Combining these programs into a parallel system (see Section~\ref{sec:network-lang}) gives an executable interpretation of the entire choreography.
Specifically, we lift EPP point-wise to a finite set of locations $\SysLocs \subseteq \Locations$, defining $\epp{C}{\SysLocs} = {\parallel_{L \in \SysLocs} (L \triangleright \epp{C}{L})}$,
which also requires $\epp{C}{L}$ to be defined for all $L \in \SysLocs$.

\subsection{Bisimulation Relation}\label{sec:bisim}
Next, we will examine the relationship between our choreographic operational semantics and the semantics given by EPP, with two main goals in mind.
Firstly, to exhibit a bisimulation between these two semantics to justify the correctness of EPP.
Secondly, to give a guarantee of deadlock freedom by design for compiled systems.

To construct a bisimulation, we must choose which systems are related to a given choreography.
The natural choice is to say that a choreography~$C$ is related only to its projection~$\epp{C}{\SysLocs}$.
However, this relation is too strict.
When a choreography branches, the branch which is not taken is discarded,
but in the projected system, that branch may be preserved by locations waiting on selection messages.
For instance, consider the following example where~$C_1$ reduces to~$C_2$.
\begin{center}
  \begin{tikzpicture}
    \node (C1) at (0,0) {
      $\begin{array}{@{}l@{}}
        \addtocounter{numlevels}{1}
        \ITEBase*{\Alice.\True}
             {\syncs{\Alice}{\Left}{\Bob} \seq \Bob.1}
             {\syncs{\Alice}{\Right}{\Bob} \seq \Bob.2}
        \addtocounter{numlevels}{-1}
      \end{array}$
    };
    \node (C2) at (5,0) {$\syncs{\Alice}{\Left}{\Bob} \seq \Bob.1$};
    \node (C1Proj) at (0,-2.2) {
      $\begin{array}{@{}l@{}}
        \addtocounter{numlevels}{1}
        \AllowChoice*{\Alice}{\Ret{1}}{\Ret{2}}
        \addtocounter{numlevels}{-1}
      \end{array}$
    };
    \node (C2Proj) at (5,-2.2) {
      $\begin{array}{@{}l@{}}
        \addtocounter{numlevels}{1}
        \AllowOneChoice*{\Alice}{\Left}{\Ret{1}}
        \addtocounter{numlevels}{-1}
      \end{array}$
    };

    \node[base left=0pt of C1,val node] {$C_1 =$};
    \node[base right=0pt of C2,val node] {$= C_2$};
    \node[left=0pt of C1Proj,val node] {$\epp{C_1}{\Bob} =$};
    \node[right=0pt of C2Proj,val node] {$= \epp{C_2}{\Bob}$};

    \draw[-implies,double equal sign distance,shorten >=2pt] (C1.east) -- (C2.west) node[pos=0.95,below]{${}_c$} node[label,midway,above]{$\RIfTrue{\Alice}$};
    \draw[mapsto] (C1) -- (C1Proj) node[label,midway,left,yshift=1.5pt]{$\epp{\cdot}{\Bob}$};
    \draw[mapsto] (C2) -- (C2Proj) node[label,midway,right,yshift=1.5pt]{$\epp{\cdot}{\Bob}$};
    \draw[red,dashed] (C1Proj.east) -- (C2Proj.west) node[label,above,midway,red]{$\neq$};
  \end{tikzpicture}
\end{center}
Once~$C_1$ takes the left branch, \Alice~can never send the selection message~\Right to~\Bob.
While this makes sense from \Alice's perspective, it does not make sense for~\Bob, whose projected program has discarded the \Right~branch without receiving any input from~\Alice.
We expect that choreographic steps will only affect the projection of locations involved in the step.

An additional wrinkle stems from EPP's use of the collapsing sequencing function~$E_1 \seqfun E_2$:
local programs (in an arbitrary subexpression) which resolve to a value after a substitution may be removed from the projected program, as in the following example.
\begin{center}
  \begin{tikzpicture}
    \node (C1) at (0,0) {
      $\begin{array}{@{}l@{}}
        \addtocounter{numlevels}{1}
        \LetIn*{\{\Alice,\Bob\}.x}{\{\Alice,\Bob\}.1}
                   {\Fun*{F}{X}{\enspace\LetIn{\Alice.y}{\{\Alice,\Bob\}.x}{\Alice.2}}}
        \addtocounter{numlevels}{-1}
      \end{array}$
    };
    \node (C2) at (7,0) {
      $\begin{array}{@{}l@{}}
        \addtocounter{numlevels}{1}
        \Fun*{F}{X}{\enspace\LetIn{\Alice.y}{\{\Alice,\Bob\}.1}{\Alice.2}}
        \addtocounter{numlevels}{-1}
      \end{array}$
    };
    \node (C1Proj) at (0,-2.2) {$\Newline*{\CtrlLetIn{x}{\Ret{1}}{}}{\CtrlFun{F}{X}{\Ret{x} \CtrlSeq \CtrlUnit}}$};
    \node (C2Proj) at (7,-2.2) {$\CtrlFun{F}{X}{\CtrlUnit}$};

    \node[base left=0pt of C1,val node] {$C_1 =$};
    \node[base right=0pt of C2,val node] {$= C_2$};
    \node[left=0pt of C1Proj,val node] {$\epp{C_1}{\Bob} =$};
    \node[right=0pt of C2Proj,val node] {$= \epp{C_2}{\Bob}$};

    \draw[-implies,double equal sign distance,shorten >=2pt] (C1.east) -- (C2.west) node[pos=0.96,below]{${}_c$} node[label,midway,above]{$\RLet{\{\Alice,\Bob\}.x}{1}$};
    \draw[mapsto] (C1) -- (C1Proj) node[label,midway,left,yshift=1.5pt]{$\epp{\cdot}{\Bob}$};
    \draw[mapsto] (C2) -- (C2Proj) node[label,midway,right,yshift=1.5pt]{$\epp{\cdot}{\Bob}$};
    \draw[-implies,double equal sign distance,red] (C1Proj.east) -- (C2Proj.west) node[label,midway]{$\diagup$};
  \end{tikzpicture}
\end{center}
This reduction substitutes~$1$ for~$x$ in the body of the function.
While~$\epp{C_1}{\Bob}$ can analogously substitute~$x$, the resultant program~$\CtrlFun{F}{X}{\Ret{1} \CtrlSeq \CtrlUnit}$ differs from $\epp{C_2}{\Bob}$,
as the projection collapses the body of the function to~$\Ret{1} \seqfun \CtrlUnit = \CtrlUnit$.
As the mismatch is in the function body, there are no steps~\Bob can take to correct it.

We account for both types of mismatches using a relation $E_1 \lessnd E_2$
indicating that~$E_1$ may have discarded some unneeded code---either choices or sequenced values---that~$E_2$ retains.
Formally, it is the smallest partial order on network programs that is structurally compatible---it admits rules like
$\frac{
  E_1 \mkern3mu\lessnd\mkern3mu E_1'
  \quad
  E_2 \mkern3mu\lessnd\mkern3mu E_2'
}{E_1 \mkern2mu\CtrlSeq\mkern2mu E_2 \mkern3mu\lessnd\mkern3mu E_1' \mkern2mu\CtrlSeq\mkern2mu E_2'}$---%
and admits the following three rules.
The first two handle additional branches, while the third covers collapsing sequences.
\begin{mathpar}[\rulefiguresize]
  \infer{E_1 \lessnd E_1'}
    {\AllowOneChoiceTight*{\ell}{\CtrlLeft}{E_1} \lessnd \AllowChoiceTight*{\ell}{E_1'}{E_2'}}
  \and
  \infer{E_2 \lessnd E_2'}
    {\AllowOneChoiceTight*{\ell}{\CtrlRight}{E_2} \lessnd \AllowChoiceTight*{\ell}{E_1'}{E_2'}}
  \and
  \infer{
    E_1 \lessnd E_2 \\
    \CtrlVal{V}
  }{E_1 \lessnd V \CtrlSeq E_2}
\end{mathpar}

The relation also lifts point-wise to systems (with identical domains $\SysLocs$):
\[
  \Pi_1 \lessnd \Pi_2 ~\defeq~ \forall L \in \SysLocs \ldotp \Pi_1(L) \lessnd \Pi_2(L).
\]

By relaxing the bisimulation to relate a choreography~$C$ not only to~$\epp{C}{\SysLocs}$,
but to any system~$\Pi$ where $\epp{C}{\SysLocs} \lessnd \Pi$, we avoid the problems described above.
In the first example, $\epp{C_2}{\Bob} \lessnd \epp{C_1}{\Bob}$,
and in the second, $\epp{C_1}{\Bob} \ctrlstep{} E$ where $\epp{C_2}{\Bob} \lessnd E$, as needed.

\subsection{Soundness, Completeness, and Deadlock Freedom}
\label{sec:proj-correct}
The above correspondence is sufficient to show that our choreographic and projected system semantics are equivalent and yield deadlock freedom of compiled systems as a result.
To this end, we prove the choreographic semantics is sound and complete with respect to the projected system.

The theorems require that all \emph{location names in~$C$}, denoted~$\namedlocs{C}$,
are included in the projected system: $\namedlocs{C} \subseteq \SysLocs$.
This requirement formalizes the implicit assumption that all locations specified by the choreography are present in the system,
lest, for instance,~\Alice be unable to compute $\Alice.2 \ColSend \Bob$ because~\Bob is not in the system.
We similarly require $\SysLocs \neq \varnothing$ to ensure \emph{someone} is computing.

The completeness theorem is straightforward, and says the projected semantics simulate the choreographic semantics.
\begin{thm}[Completeness]\label{thm:completeness}
  If $\chortyped{\Gamma}{\Delta}{\Sigma}{C}{\tau}$, $C \stepsn{n} C'$, and $\namedlocs{C} \subseteq \SysLocs \neq \varnothing$,
  then there is some $\Pi$ and $k \geq n$ such that $\epp{C}{\SysLocs} \systemstepsn{k} \Pi$ and $\epp{C'}{\SysLocs} \lessnd \Pi$.
\end{thm}
The mechanized proof first shows the system can simulate a single step and then extends it to multiple steps by induction.
It also uses the fact that, if~$E_1 \lessnd E_2$ and $E_1$ is the projection of a choreography,
then~$E_1$ and~$E_2$ can make the same reductions, plus $E_2$ possibly taking administrative steps to remove extra un-collapsed semicolons.

We would like a soundness theorem that says the choreographic semantics can simulate the projected system,
but there are some technical complications.
First, after a single reduction in a projected system, the reduced system does not necessarily correspond to any choreography, and may instead be some intermediate state.
The following example shows a case where the projected system~$\epp{C_1}{\{\Alice,\Bob\}}$ can reduce to a state~$\Pi$ which no choreography projects to.
\begin{center}
  \newlength{\vertgap}
  \setlength{\vertgap}{3em}
  \begin{tikzpicture}
    \node (C1Proj) {$\Newline*{\Alice \triangleright \Ret{2 + 3}}{\Bob \triangleright \Ret{2 + 3}}$};
    \node[above=\vertgap of C1Proj] (C1) {$\{\Alice,\Bob\}.(2 + 3)$};
    \node[right=2.5em of C1Proj] (Middle) {$\Newline*{\Alice \triangleright \Ret{5}}{\Bob \triangleright \Ret{2 + 3}}$};
    \node[above=0pt of Middle,val node,anchor=west,rotate=90] (eq) {$=$};
    \node[above=0pt of eq.east] (Pi) {$\Pi$};
    \node[right=2.5em of Middle] (C2Proj) {$\Newline*{\Alice \triangleright \Ret{5}}{\Bob \triangleright \Ret{5}}$};
    \node[above=\vertgap of C2Proj] (C2) {$\{\Alice,\Bob\}.5$};

    \node[base left=0pt of C1,val node] {$C_1 =$};
    \node[base right=0pt of C2,val node] {$= C_2$};
    \node[left=0pt of C1Proj,val node] {$\epp{C_1}{\{\Alice,\Bob\}} =$};
    \node[right=0pt of C2Proj,val node] {$= \epp{C_2}{\{\Alice,\Bob\}}$};

    \draw[-implies,double equal sign distance,shorten >=2pt] (C1.east) -- (C2.west) node[pos=0.99,below]{${}_c$};
    \draw[-implies,double equal sign distance,shorten >=2pt] (C1Proj.east) -- (Middle.west) node[pos=0.94,below]{${}_S$};
    \draw[-implies,double equal sign distance,shorten >=2pt] (Middle.east) -- (C2Proj.west) node[pos=0.94,below]{${}_S$};
    \draw[mapsto] (C1) -- (C1Proj) node[label,midway,left,yshift=1.5pt]{$\epp{\cdot}{\{\Alice,\Bob\}}$};
    \draw[mapsto] (C2) -- (C2Proj) node[label,midway,right,yshift=1.5pt]{$\epp{\cdot}{\{\Alice,\Bob\}}$};
  \end{tikzpicture}
\end{center}
To reach the system corresponding to the projection of~$C_2$, $\Pi$ must continue to progress.
As a result, even our single-step soundness theorem must allow the system to continue to reduce
until it reaches a point where the original choreography can match its progress.

The second technicality is a fundamental challenge stemming from the combination of nontermination and multiply-located local computations.
In the system, each participant in a multiply-located local computation evaluates it independently.
If one location performs such a step while another is blocked by an earlier infinite loop, however, the choreography will be unable to match the system.
For instance, in $\LetIn{\Alice.x}{\Alice.\Loop}{\{\Alice,\Bob\}.(1 \LocalPlus 1)}$ the only possible choreographic step is to run the infinite~\Loop at~\Alice,
but in the projected system, \Bob~may reduce the local expression $1 \LocalPlus 1$ to $2$.

To avoid such situations, our soundness theorem only holds for choreographies~$C$ where all local computations terminate.
Specifically, if~$C \steps{} C'$ and~$\rho.e$ is in an evaluation position of~$C'$, then~$e$ must terminate.
Importantly, this condition does not require the local \emph{language} to be terminating, only the specific local computations that execute in~$C$.

With these two caveats in place, we can now state a single-step version of soundness.

\begin{prop}[Single-Step Soundness]\label{prop:single-step-soundness}
  If $\choremptyped{C}{\tau}$, $\epp{C}{\SysLocs} \systemstep \Pi$, $\namedlocs{C} \subseteq \SysLocs \neq \varnothing$,
  and all local programs that~$C$ executes terminate, then for some $\Pi'$ and $C'$, $\Pi \systemsteps \Pi'$, $C \stepss C'$, and $\epp{C'}{\SysLocs} \lessnd \Pi'$.
\end{prop}

Extending this result to simulate multiple system steps with a na\"{i}ve induction runs into another problem:
the steps that we wish to simulate and the extra---possibly different---steps needed to ensure the system is not in an intermediate state
are independent reduction sequences. 
To reduce the endpoints of these diverging paths to a common system, we employ a confluence theorem, which critically relies on confluence of the local language.
\begin{prop}[System Confluence]\label{prop:confluence}
  If $\Pi_1 \systemsteps \Pi_2$ and $\Pi_1 \systemsteps \Pi_3$, then there is some $\Pi_4$ such that $\Pi_2 \systemsteps \Pi_4$ and $\Pi_3 \systemsteps \Pi_4$.
\end{prop}

Combining confluence with Proposition~\ref{prop:single-step-soundness}
is enough to prove soundness beginning with any number of system steps.
\begin{thm}[Soundness]\label{thm:soundness}
  If $\choremptyped{C}{\tau}$, $\epp{C}{\SysLocs} \systemsteps \Pi$, $\namedlocs{C} \subseteq \SysLocs \neq \varnothing$,
  and all local programs that~$C$ executes terminate, then for some $\Pi'$ and $C'$, $\Pi \systemsteps \Pi'$, $C \steps{} C'$, and $\epp{C'}{\SysLocs} \lessnd \Pi'$.
\end{thm}

Having demonstrated the correspondence between the choreographic and system semantics, we can now prove deadlock-freedom.
Although bisimulation only holds when all executed local programs terminate, deadlock-freedom holds for all projected systems, even in the presence of non-terminating local computations.
By combining the soundness of the type system (Theorems~\ref{thm:preservation} and~\ref{thm:progress}) and the completeness of EPP (Theorem~\ref{thm:completeness}), we prove that the system either enters an infinite loop for some location or terminates in a value for all locations, which is sufficient.
\begin{thm}[Deadlock Freedom by Design]\label{thm:deadlock-freedom}
  If $\choremptyped{C}{\tau}$, $\epp{C}{\SysLocs} \systemsteps \Pi$, and $\namedlocs{C} \subseteq \SysLocs$,
  then either every location in~$\Pi$ maps to a value, or there is some~$\Pi'$ such that $\Pi \systemstep \Pi'$.
\end{thm}



\section{Rocq Development}
\label{sec:rocq-dev}
We now discuss some details of our Rocq formalization and how they differ from the presentation above.
For clarity, we presented \langname in a standard style with named variables in the paper,
but for ease of development, the Rocq code uses a nameless style with de~Bruijn indices for both program and type variables.
As in Pirouette~\cite{HirschG22}, the local language must also use de~Bruijn indices, and provide common guarantees about substitution.

This change also forces a different handling of location- and location-set--variable substitution in the Rocq code.
As described in Section~\ref{sec:location-substitution}, care must be taken when performing location substitution to avoid variable capture between the namespaces of different locations.
The Rocq encoding, however, avoids this issue by not separating namespaces by location.
Instead, it treats all local variables uniformly as part of a single global namespace of de~Bruijn indices.
With this formulation, no capture is possible upon substituting location variables when using the standard capture-avoiding definition of location substitution.

Finally, our formalization proves deadlock freedom in two pieces.
The first uses bisimulation (Theorems~\ref{thm:completeness} and~\ref{thm:soundness} together) to prove deadlock-freedom when all local programs that execute terminate.
The second assumes the presence of a non-terminating local computation and proves deadlock freedom directly.
The proof of Theorem~\ref{thm:deadlock-freedom}, which does not condition on local (non\nobreakdash-)termination
is immediate by combining these, but assumes the law of the excluded middle~(LEM) to differentiate between the cases.
We make the LEM a premise to this theorem to explicitly indicate the use of a non-constructive assumption.


\section{Related Work}
\label{sec:related-work}

As mentioned in Section~\ref{sec:introduction}, choreographic programming has seen substantial recent development.
This paper fits into the emerging paradigm of \emph{functional} choreographic programming
with process polymorphism, originally proposed by~\citet{GraversenHM24}.
Choreographic programming as a whole arises from concurrency theory and the study of $\pi$-calculi.
We discuss each of these in turn.

\subsection{Functional Choreographic Programming}
\label{sec:choreo-prog}

Choreographic programming was crystallized as an independent paradigm in 2013 by the work of \citet{CarboneM13}, especially Montesi's~Ph.D. thesis~\citep{Montesi13}.
For the next 10 years, it advanced, but remained tied to a lower-order model of computing \cite[see, e.g.,][]{CarboneMS14,Cruz-FilipeMP18,Cruz-FilipeM17,LaneseMZ13,Cruz-FilipeM17c}.

That bind was broken in 2022 by two independent developments: Pirouette~\cite{HirschG22} and Chor$\lambda$~\cite{CruzFilipeGLMP22}.
These works combine the primitives of choreographic programming with $\lambda$~calculi, allowing for (sequential) composition of choreographies and program reuse.
However, neither support any type of polymorphism or multiply-located values.

\langname is based on Pirouette, inheriting many of its features including
a separate language of messages (the local language)
and out-of-order semantics that require all participants to synchronize on choreographic function applications.
Chor$\lambda$, by contrast, combines the languages of choreographies and messages.
It originally had a strictly sequential semantics~\cite{CruzFilipeGLMP22},
but recent work showed how to give it an out-of-order semantics \emph{without} global synchronization~\citep{CruzFilipeGLMP23}.
These semantics use commuting conversions, a set of semantics-preserving rewrite rules
such as $(\lambda X\ldotp C_1)~C_2~C_3 \Rrightarrow (\lambda X\ldotp C_1~C_3)~C_2$, and allow reductions inside function bodies.
However, commuting conversions are fragile.
For instance, they fail to preserve types in the presence of named recursive functions, like those in \langname.
That is, the hypothetical rewrite $(\Fun{F}{X}{C_1})~C_2~C_3 \Rrightarrow (\Fun{F}{X}{C_1~C_3})~C_2$ changes the type of~$F$, which is bound in~$C_1$.
Polymorphism and recursive types produce similar difficulties.

We thus adopt the more restrictive semantics of Pirouette,
and leave finding an appropriate semantics without global synchronization as future work.
Importantly, nothing in this paper relies on this global synchronization except for the \emph{statement} of EPP correctness (Section~\ref{sec:proj-correct}).
We believe an appropriate statement of EPP correctness could support a semantics without this synchronization,
but identifying such a statement requires significant research in its own right.

\subsection{Process Polymorphism}
\label{sec:process-poly}

More recently, \citet{GraversenHM24} extended Chor$\lambda$ with process polymorphism,
allowing choreographies to abstract over (the identity of) their participant processes.
This extension, PolyChor$\lambda$, introduced the process abstraction and shows how to project it by adding the ``AmI'' construct to the target language.
PolyChor$\lambda$ is based on System $F_\omega$, and includes quantification over processes, types, and types of higher-order kinds, but does not include quantification over \emph{sets} of processes.
Their work also does not attempt to define an out-of-order semantics, and instead settles for a sequential, call-by-value semantics for its choreographies.

\citet{BatesKJSKN25} separately introduced the idea of \emph{census polymorphism}, which allows choreographies to abstract over the quantity and identity of their participants.
Their work presents extensive real-world uses cases for multiply-located values and abstraction over sets of processes
and highlights several practical implementations (e.g., in MultiChor, ChoRus, and ChoreoTS).
However, the formal model is limited and lacks deadlock-freedom guarantees in the presence of polymorphism.

Our work not only supports first-class location set polymorphism and multiply-located values (missing from PolyChor$\lambda$)
and a deadlock freedom guarantee (missing from the work of \citeauthor{BatesKJSKN25}),
but we mechanically verify all results in Rocq.
To our knowledge, the only prior mechanizations of choreographic results concern Pirouette~\citep{HirschG22}
and simpler lower-order systems~\citep{CruzFilipeMP21a,CruzFilipeMP21b},
meaning this work contains the first formalization of choreographic process polymorphism or of multiply-located values,
let alone \langname's full suite of features.

\subsection{Higher-Order Communication}
\label{sec:higher-order-comm}
Another important aspect of concurrent systems is \emph{higher-order communication}:
communicating a channel (or channel name) over a channel.
This feature is critical to model systems with a dynamic communication topology, and can be expressed natively in many untyped process calculi such as the $\pi$-calculus and its higher-order variant the $\text{HO}\pi$-calculus~\cite{Sangiorgi93}.
Previous work proving deadlock-freedom for concurrent languages with higher-order communication
has relied on advanced typing regimes, like higher-order session types~\cite[see, e.g.,][]{MostrousY07,CostaMPV22,PocasCMV23},
that are highly complex and impose significant restrictions on communication patterns, like requiring them to be from a regular or context-free grammar.
There exist simpler (multiparty) session type systems based on linear and separation logic that support first-class--sessions and --endpoints~\citep{JacobsBK22,JacobsHK24},
but they must impose an acyclic inter-session communication topology to guarantee deadlock freedom.

PolyChor$\lambda$ supports a restricted form of higher-order communication using \emph{delegation}, where one process can request that another perform an interaction on their behalf.
In particular, they allow processes to communicate entire choreographies, which may themselves contain communication.
However, the type system prevents delegated computation from containing unresolved location variables, limiting the expressive power.

Computing and sending first-class location names is a distinct form of higher-order communication
from both $\pi$-calculus channels and PolyChor$\lambda$'s delegation.
Channels represent a limited capacity to perform a specific action, like sending one value,
while choreographic process names are identifiers of participants who may be asked, by name, to perform many actions.
Meanwhile, providing first-class data and location names to choreographic functions and type abstractions
replicate much of the functionality of delegation, but must be specified manually.


\section{Conclusion}
\label{sec:conclusion}

This work presented the choreographic Quick Change calculus, \langname, a choreographic calculus supporting process (set) and type polymorphism, algebraic and recursive data types, multiply-located values, and first-class location names.
While prior choreographic languages implement some of the first three features, none support all three,
and \langname is the first to support first-class location names in a typed choreography.

We showed how to integrate polymorphism over types, processes, and sets of processes using a System F-like type system.
The \langname type system simultaneously allows first-class treatment of (sets of) location names and considers locations type-level values.
Yet it still avoids dependency by restricting the types of computations which use these locations to not depend on values.
To prevent deadlocks, \langname statically ensures that each location knows if it has been selected by any dynamic location computation.

We also showed how to project the constructs in \langname to a network language
that faithfully models the concurrent execution of a multi-party system,
and we proved the classic choreographic result:
all systems projected from well-typed choreographies are deadlock-free.
We mechanically verified proofs of this result as well as others in the paper,
providing the first mechanized formalization for any choreography supporting any form of process polymorphism or multiply-located values.

While there is still a wealth of future research in expanding and refining the features of our calculus, even now, it is able to facilitate practical and safe programming of concurrent systems.
Our paradigmatic example of a distributed thread pool with a load balancer
can be concisely written in \langname, yet is far beyond the capabilities of
prior typed choreographic languages.


\section*{Acknowledgments}
Thanks to the anonymous reviewers for their valuable feedback and thoughtful suggestions.
Also thanks to Rahul Krishnan for help editing, and to Keith Allen for useful comments on an earlier draft.
Support for this research was provided by the University of Wisconsin--Madison
Office of the Vice Chancellor for Research with funding from the Wisconsin Alumni Research Foundation.

\bibliography{../bibliography/main,tr}

\appendix
\section*{Appendices}
\section{Choreography Operational Semantics}
\subsection{Choreography Values}
\begin{syntax}
  \category[Choreography Values]{V}
  \alternative{\rho.v} \alternative{\Fun{F}{X}{C}} \alternative{\TFun{\alpha \knd \kappa}{C}}\\
  \alternative{(V_1,V_2)} \alternative{\Inl{V}} \alternative{\Inr{V}} \alternative{\Fold{V}}
\end{syntax}

\subsection{Redices and Evaluation Contexts}
\begin{syntax}
  \category[Messages]{m}
  \alternative{v} \alternative{d}

  \category[Redices]{R}
  \alternative{\RDone{\rho}{e_1}{e_2}}
  \alternative{\RFun{R}}
  \alternative{\RArg{R}}
  \alternative{\RApp}
  \alternative{\RTApp} 
  \alternative{\RUnfoldFold} \\
  \alternative{\RPairL{R}}
  \alternative{\RPairR{R}}
  \alternative{\RFstPair}
  \alternative{\RSndPair}
  \alternative{\RCaseInl} 
  \alternative{\RCaseInr} \\
  \alternative{\RLet{\rho}{v}}
  \alternative{\RLet{\rho}{t}}
  \alternative{\RSendV{\ell}{m}{\rho}}
  \alternative{\RIfTrue{\rho}}
  \alternative{\RIfFalse{\rho}}\\

  \category[Evaluation Contexts]{\eta}
  \alternative{\hole~C}
  \alternative{V~\hole}
  \alternative{\hole~t}
  \alternative{\Fold{\hole}}
  \alternative{\Unfold{\hole}}\\
  \alternative{(\hole,C)}
  \alternative{(V,\hole)}
  \alternative{\Fst{\hole}}
  \alternative{\Snd{\hole}} \\
  \alternative{\Inl{\hole}}
  \alternative{\Inr{\hole}}
  \alternative{\Case{\hole}{X}{C_1}{Y}{C_2}}\\
  \alternative{\LetIn{\rho.x \ty t_e}{\hole}{C_2}}
  \alternative{\LetIn{\rho.\alpha \knd \kappa}{\hole}{C_2}}\\
  \alternative{\hole \ChorSend[\ell] \rho}
  \alternative{\ITE{\hole}{C_1}{C_2}}
\end{syntax}

\subsection{Redex Blocked Locations}
\begin{mathparpagebreakable}
  \rloc{\RDone{\rho}{e_1}{e_2}} = \rho \and
  \rloc{\RFun{R}} = \rloc{R} \and
  \rloc{\RArg{R}} = \rloc{R} \and
  \rloc{\RApp} = \Locations \and
  \rloc{\RTApp} = \Locations \and
  \rloc{\RUnfoldFold} = \Locations \and
  \rloc{\RPairL{R}} = \rloc{R} \and
  \rloc{\RPairR{R}} = \rloc{R} \and
  \rloc{\RFstPair} = \Locations \and
  \rloc{\RSndPair} = \Locations \and
  \rloc{\RCaseInl} = \Locations \and
  \rloc{\RCaseInr} = \Locations \and
  \rloc{\RLet{\rho}{v}} = \rho \and
  \rloc{\RLet{\rho}{t}} = \rho \and
  \rloc{\RSendV{\ell}{m}{\rho}} = \{\ell\} \cup \rho \and
  \rloc{\RIfTrue{\rho}} = \rho \and
  \rloc{\RIfFalse{\rho}} = \rho
\end{mathparpagebreakable}

\subsection{Choreography Blocked Locations}
\begin{mathparpagebreakable}
  \cloc{X} = \varnothing
  \and
  \cloc{\rho.e} = \rho
  \and
  \cloc{\Fun{F}{X \ty \tau}{C}} = \varnothing
  \and
  \cloc{C_1~C_2} = \Locations
  \and
  \cloc{\TFun{\alpha \knd \kappa}{C}} = \varnothing
  \and
  \cloc{C~t} = \Locations
  \and
  \cloc{\Fold{C}} = \cloc{C}
  \and
  \cloc{\Unfold{C}} = \Locations
  \and
  \cloc{(C_1,C_2)} = \cloc{C_1} \cup \cloc{C_2}
  \and
  \cloc{\Fst{C}} = \Locations
  \and
  \cloc{\Snd{C}} = \Locations
  \and
  \cloc{\Inl{C}} = \cloc{C}
  \and
  \cloc{\Inr{C}} = \cloc{C}
  \and
  \cloc{\Case{C}{X}{C_1}{Y}{C_2}} = \Locations
  \and
  \cloc{\LetIn{\rho.x \ty t_e}{C_1}{C_2}} = \rho \cup \cloc{C_1} \cup \cloc{C_2}
  \and
  \cloc{\LetIn{\rho.\alpha \knd \kappa}{C_1}{C_2}} = \rho \cup \cloc{C_1} \cup \cloc{C_2}
  \and
  \cloc{C \ChorSend[\ell] \rho} = \{\ell\} \cup \rho \cup \cloc{C}
  \and
  \cloc{\syncs{\ell}{d}{\rho} \seq C} = \{\ell\} \cup \rho \cup \cloc{C}
  \and
  \cloc{\ITE{C}{C_1}{C_2}} = \rho \cup \cloc{C} \cup \cloc{C_1} \cup \cloc{C_2}
\end{mathparpagebreakable}

\subsection{Redex for an Evaluation Context}
If $\eta$ is an evaluation context and $R$ is a redex, we define $\eta[R]$ to be the redex which corresponds to making the reduction given by $R$ in the context $\eta$.
\begin{mathparpagebreakable}
    \ctxredex{(\hole~C)} \defeq \RFun{R}
    \and
    \ctxredex{(V~\hole)} \defeq \RArg{R}
    \and
    \ctxredex{(\hole~t)} \defeq R
    \and
    \ctxredex{(\Fold{\hole})} \defeq R
    \and
    \ctxredex{(\Unfold{\hole})} \defeq R
    \and
    \ctxredex{(\hole,C)} \defeq \RPairL{R}
    \and
    \ctxredex{(V,\hole)} \defeq \RPairR{R}
    \and
    \ctxredex{(\Fst{\hole})} \defeq R
    \and
    \ctxredex{(\Snd{\hole})} \defeq R
    \and
    \ctxredex{(\Inl{\hole})} \defeq R
    \and
    \ctxredex{(\Inr{\hole})} \defeq R
    \and
    \ctxredex{(\Case{\hole}{X}{C_1}{Y}{C_2})} \defeq R
    \and
    \ctxredex{(\hole \ChorSend[\ell] \rho)} \defeq R
    \and
    \ctxredex{(\LetIn{\rho.x \ty t_e}{\hole}{C})} \defeq R
    \and
    \ctxredex{(\LetIn{\alpha \knd \kappa}{\hole}{C})} \defeq R
    \and
    \ctxredex{(\ITE{\hole}{C_1}{C_2})} \defeq R
\end{mathparpagebreakable}

\subsection{Location Set Relations}\label{sec:set-relations}
Here we define precisely the containment~$\ell \in \rho$, disjointness~$\rho_1 \cap \rho_2 = \varnothing$, and subset $\rho_1 \subseteq \rho_2$ relations, with special care given to how they are defined when the types in question are non-ground.
In particular, we define two versions of containment: \emph{necessary} containment~$\Nec (\ell \in \rho)$, and \emph{possible} containment~$\Pos (\ell \in \rho)$.
The un-annotated containment relation is treated as necessary containment ($\ell \in \rho \defeq \Nec(\ell \in \rho)$), and the disjointness relation (to be interpreted as necessary disjointness) is defined as follows:
\[(\rho_1 \cap \rho_2 = \varnothing) \defeq \forall \ell \ldotp \neg (\Pos (\ell \in \rho_1) \wedge \Pos (\ell \in \rho_2)). \]
The difference between the two versions of containment is that location sets which are variables (or a singleton of a variable) can possibly contain any location, while variable location sets do not necessarily contain any location.
Note here that the metavariable~$\ell$ stands for either a type variable~$\alpha$ or a concrete location~$L \in \Locations$, and the metavariable~$\rho$ stands for any location set, including possibly a type variable.
\begin{mathparpagebreakable}
  \infer{~}{\Nec (\ell \in \{\ell\})}
  \and
  \infer{\Nec (\ell \in \rho_1)}{\Nec (\ell \in \rho_1 \cup \rho_2)}
  \and
  \infer{\Nec (\ell \in \rho_2)}{\Nec (\ell \in \rho_1 \cup \rho_2)}
  \\
  \infer{~}{\Pos (\ell \in \alpha)}
  \and
  \infer{~}{\Pos (\ell \in \{\alpha\})}
  \and
  \infer{~}{\Pos (L \in \{L\})}
  \and
  \infer{\Pos (\ell \in \rho_1)}{\Pos (\ell \in \rho_1 \cup \rho_2)}
  \and
  \infer{\Pos (\ell \in \rho_2)}{\Pos (\ell \in \rho_1 \cup \rho_2)}
\end{mathparpagebreakable}

To define the (necessarily a) subset relation, we first note that we cannot use the na\"{i}ve definition in terms of the necessary containment relation.
That is, $\forall \ell \ldotp \Nec (\ell \in \rho_1) \Rightarrow \Nec (\ell \in \rho_2)$ would not serve as a correct definition for $\rho_1 \subseteq \rho_2$ in the presence of type variables.
This is because~$\neg \Nec (\ell \in \alpha)$ for every location~$\ell$, so with this definition we would have that~$\alpha \subseteq \rho$ for every set~$\rho$.
The subset relation (and relations of~$\Nec$ modality in general) should be preserved under substitution, but our example shows that this is not the case when using the na\"{i}ve definition.
Instead, the subset relation must be defined inductively as follows.
\begin{mathparpagebreakable}
  \infer{~}{\varnothing \subseteq \rho}
  \and
  \infer{~}{\rho \subseteq \rho}
  \and
  \infer{\rho \subseteq \rho_1}{\rho \subseteq \rho_1 \cup \rho_2}
  \and
  \infer{\rho \subseteq \rho_2}{\rho \subseteq \rho_1 \cup \rho_2}
  \and
  \infer{\Nec (\ell \in \rho)}{\{\ell\} \subseteq \rho}
  \and
  \infer{\rho_1 \subseteq \rho \\
  \rho_2 \subseteq \rho}
  {\rho_1 \cup \rho_2 \subseteq \rho}
\end{mathparpagebreakable}

\subsection{Choreography Operational Semantics}
\label{sec:full-chor-sem}
\begin{mathparpagebreakable}
  \CCtxRule
  \and
  \CDoneRule
  \and
  \CAppRule
  \and
  \CTAppRule
  \and
  \CUnfoldFoldRule
  \and
  \CFstPairRule
  \and
  \CSndPairRule
  \and
  \CCaseInlRule
  \and
  \CCaseInrRule
  \and
  \CLetVRule
  \and
  \CLetIRule
  \and
  \CTyLetVRule
  \and
  \CTyLetIRule
  \and
  \CSendVRule
  \and
  \CSyncRule
  \and
  \CSyncIRule
  \and
  \CIfTRule
  \and
  \CIfFRule
  \and
  \CIfIRule
  \and
  \CAppIRule
  \and
  \CPairIRule
\end{mathparpagebreakable}


\section{Static Semantics}
\subsection{\langname Kinding System}
\label{sec:full_kinds}
\begin{mathparpagebreakable}[\rulefiguresize]
  \KVarRule
  \and
  \KLocalRule
  \and
  \KAtRule
  \and
  \KArrowRule
  \and
  \KAllRule
  \and
  \KProdRule
  \and
  \KSumRule
  \and
  \KRecRule
  \and
  \KLocRule
  \and
  \KSngRule
  \and
  \KUnionRule
  \and
  \WFEmpCtxRule
  \and
  \WFAddCtxRule
  \and
  \WFAddLocalCtxRule
\end{mathparpagebreakable}

\subsection{\langname Type System}
\label{sec:full_chor_types}
\begin{mathparpagebreakable}[\rulefiguresize]
  \TVarRule
  \and
  \TDoneRule
  \and
  \TFunRule
  \and
  \TAppRule
  \and
  \TAbsRule
  \and
  \TTAppRule
  \and
  \TFoldRule
  \and
  \TUnfoldRule
  \and
  \TPairRule
  \and
  \TFstRule
  \and
  \TSndRule
  \and
  \TInlRule
  \and
  \TInrRule
  \and
  \TCaseRule
  \and
  \TLetLocalRule
  \and
  \TLetLocRule
  \and
  \TLetLocSetRule
  \and
  \TLetLocalTyRule
  \and
  \TSendRule
  \and
  \TSyncRule
  \and
  \TIfRule
\end{mathparpagebreakable}


\section{Network Language}
\subsection{Network Language Expressions}\label{sec:full-ntwk-syntax}
\begin{syntax}
  \category[Network Program]{E}
  \alternative{X} \alternative{\CtrlUnit} \alternative{\Ret{e}} \alternative{E_1 \CtrlSeq E_2} \\
  \alternative{\CtrlFun{F}{X}{E}} \alternative{E_1~E_2} \alternative{\CtrlTFun{\alpha}{E}} \alternative{E~t} \\
  \alternative{(E_1,E_2)} \alternative{\CtrlFst{E}} \alternative{\CtrlSnd{E}}\\
  \alternative{\CtrlInl{E}} \alternative{\CtrlInr{E}} \alternative{\CtrlCase{E}{X}{E_1}{Y}{E_2}}\\
  \alternative{\CtrlFold{E}} \alternative{\CtrlUnfold{E}}\\
  \alternative{\SendTo{E}{\rho}} \alternative{\RecvFrom{\ell}} \\
  \alternative{\CtrlLetIn{x}{E_1}{E_2}} \alternative{\CtrlLetIn{\alpha \knd \kappa}{E_1}{E_2}} \\
  \alternative{\CtrlITE{E}{E_1}{E_2}} \\
  \alternative{\ChooseFor{d}{\ell}{E}} \\
  \alternative{\AllowChoice{\ell}{{E_1}_\bot}{{E_2}_\bot}} \\
  \alternative{\AmIIn{\rho}{E_1}{E_2}}

  \category[Network Values]{V}
  \alternative{\CtrlUnit} \alternative{\Ret{v}} \alternative{\CtrlFun{F}{X}{E}}
  \alternative{\CtrlTFun{\alpha}{E}}\\
  \alternative{(V_1,V_2)} \alternative{\CtrlInl{V}} \alternative{\CtrlInr{V}} \alternative{\CtrlFold{V}}
\end{syntax}


\subsection{Transition Labels and Evaluation Contexts}
\begin{syntax}
  \category[Transition Labels]{l}
  \alternative{\iota}
  \alternative{\iotasync}
  \alternative{\RSendNtwk{m}{\rho}}
  \alternative{\RRecvNtwk{L}{m}}
  \\
  \category[Evaluation Contexts]{\eta}
  \alternative{\hole \CtrlSeq E}
  \alternative{\hole~E}
  \alternative{V~\hole}
  \alternative{\hole~t}
  \alternative{\CtrlFold{\hole}}
  \alternative{\CtrlUnfold{\hole}}\\
  \alternative{(\hole,E)}
  \alternative{(V,\hole)}
  \alternative{\CtrlFst{\hole}}
  \alternative{\CtrlSnd{\hole}}
  \alternative{\CtrlInl{\hole}}
  \alternative{\CtrlInr{\hole}}\\
  \alternative{\CtrlCase{\hole}{X}{E_1}{Y}{E_2}}\\
  \alternative{\SendTo{\hole}{\rho}}
  \alternative{\CtrlLetIn{x}{\hole}{E}}
  \alternative{\CtrlLetIn{\alpha \knd \kappa}{\hole}{E}}\\
  \alternative{\CtrlITE{\hole}{E_1}{E_2}}
\end{syntax}

\subsection{Network Language Operational Semantics}
\label{sec:full-ntwk-sem}
\begin{mathparpagebreakable}[\rulefiguresize]
  \NCtxRule
  \and
  \NRetRule
  \and
  \NSeqRule
  \and
  \NAppRule
  \and
  \NTAppRule
  \and
  \NUnfoldFoldRule
  \and
  \NFstPairRule
  \and
  \NSndPairRule
  \and
  \NCaseInlRule
  \and
  \NCaseInrRule
  \and
  \NLetRule
  \and
  \NTyLetRule
  \and
  \NSendRule
  \and
  \NRecvRule
  \and
  \NChooseRule
  \and
  \NAllowLRule
  \and
  \NAllowRRule
  \and
  \NIAmInRule
  \and
  \NIAmNotInRule
\end{mathparpagebreakable}


\section{Compilation}
\subsection{Network Program Merging}
\label{sec:ntwk-merge-def}
We show the patterns for which $E_1 \ctrlmerge E_2$ is defined; if there is no matching pattern, then $E_1 \ctrlmerge E_2$ is undefined.
\begingroup
\allowdisplaybreaks
\rulefiguresize
\begin{align*}
  \CtrlNone \ctrlmerge \CtrlNone &\defeq \CtrlNone
  \\[\vertrulegap]
  \CtrlNone \ctrlmerge E_2 &\defeq E_2
  \\[\vertrulegap]
  E_1 \ctrlmerge \CtrlNone &\defeq E_1
  \\[\vertrulegap]
  X \ctrlmerge X &\defeq X
  \\[\vertrulegap]
  \CtrlUnit \ctrlmerge \CtrlUnit &\defeq \CtrlUnit
  \\[\vertrulegap]
  \Ret{e} \ctrlmerge \Ret{e} &\defeq \Ret{e}
  \\[\vertrulegap]
  (E_{1,1} \CtrlSeq E_{1,2}) \ctrlmerge (E_{2,1} \CtrlSeq E_{2,2}) &\defeq E_{1,1} \ctrlmerge E_{2,1} \CtrlSeq E_{1,2} \ctrlmerge E_{2,2}
  \\[\vertrulegap]
  (\CtrlFun{F}{X}{E_1}) \ctrlmerge (\CtrlFun{F}{X}{E_2}) &\defeq \CtrlFun{F}{X}{(E_1 \ctrlmerge E_2)}
  \\[\vertrulegap]
  (E_{1,1}~E_{1,2}) \ctrlmerge (E_{2,1}~E_{2,2}) &\defeq (E_{1,1} \ctrlmerge E_{2,1})~(E_{1,2} \ctrlmerge E_{2,2})
  \\[\vertrulegap]
  (\CtrlTFun{\alpha \knd \kappa}{E_1}) \ctrlmerge (\CtrlTFun{\alpha \knd \kappa}{E_2}) &\defeq \CtrlTFun{\alpha \knd \kappa}{(E_1 \ctrlmerge E_2)}
  \\[\vertrulegap]
  (E_1~t) \ctrlmerge (E_2~t) &\defeq (E_1 \ctrlmerge E_2)~t
  \\[\vertrulegap]
  (\CtrlFold{E_1}) \ctrlmerge (\CtrlFold{E_2}) &\defeq \CtrlFold{(E_1 \ctrlmerge E_2)}
  \\[\vertrulegap]
  (\CtrlUnfold{E_1}) \ctrlmerge (\CtrlUnfold{E_2}) &\defeq \CtrlUnfold{(E_1 \ctrlmerge E_2)}
  \\[\vertrulegap]
  (E_{1,1},E_{1,2}) \ctrlmerge (E_{2,1},E_{2,2}) &\defeq ((E_{1,1} \ctrlmerge E_{2,1}),(E_{1,2} \ctrlmerge E_{2,2}))
  \\[\vertrulegap]
  (\CtrlFst{E_1}) \ctrlmerge (\CtrlFst{E_2}) &\defeq \CtrlFst{(E_1 \ctrlmerge E_2)}
  \\[\vertrulegap]
  (\CtrlSnd{E_1}) \ctrlmerge (\CtrlSnd{E_2}) &\defeq \CtrlSnd{(E_1 \ctrlmerge E_2)}
  \\[\vertrulegap]
  (\CtrlInl{E_1}) \ctrlmerge (\CtrlInl{E_2}) &\defeq \CtrlInl{(E_1 \ctrlmerge E_2)}
  \\[\vertrulegap]
  (\CtrlInr{E_1}) \ctrlmerge (\CtrlInr{E_2}) &\defeq \CtrlInr{(E_1 \ctrlmerge E_2)}
  \\[\vertrulegap]
  \left(\CtrlCase*{E_{1,1}}{X}{E_{1,2}}{Y}{E_{1,3}}\right) \ctrlmerge \left(\CtrlCase*{E_{2,1}}{X}{E_{2,2}}{Y}{E_{2,3}}\right) &\defeq \CtrlCase*{(E_{1,1} \ctrlmerge E_{2,1})}{X}{E_{1,2} \ctrlmerge E_{2,2}}{Y}{E_{1,3} \ctrlmerge E_{2,3}}
  \\[\vertrulegap]
  (\CtrlLetIn{x}{E_{1,1}}{E_{1,2}}) \ctrlmerge (\CtrlLetIn{x}{E_{2,1}}{E_{2,2}}) &\defeq \CtrlLetIn{x}{(E_{1,1} \ctrlmerge E_{2,1})}{(E_{1,2} \ctrlmerge E_{2,2})}
  \\[\vertrulegap]
  (\CtrlLetIn{\alpha \knd \kappa}{E_{1,1}}{E_{1,2}}) \ctrlmerge (\CtrlLetIn{\alpha \knd \kappa}{E_{2,1}}{E_{2,2}}) &\defeq \CtrlLetIn{\alpha \knd \kappa}{(E_{1,1} \ctrlmerge E_{2,1})}{(E_{1,2} \ctrlmerge E_{2,2})}
  \\[\vertrulegap]
  (\SendTo{E_1}{\rho}) \ctrlmerge (\SendTo{E_2}{\rho}) &\defeq \SendTo{(E_1 \ctrlmerge E_2)}{\rho}
  \\[\vertrulegap]
  (\RecvFrom{\ell}) \ctrlmerge (\RecvFrom{\ell}) &\defeq \RecvFrom{\ell}
  \\[\vertrulegap]
  (\ChooseFor{d}{\ell}{E_1}) \ctrlmerge (\ChooseFor{d}{\ell}{E_2}) &\defeq \ChooseFor{d}{\ell}{(E_1 \ctrlmerge E_2)}
  \\[\vertrulegap]
  \left(\AllowChoice*{\ell}{E_{1,1}}{E_{1,2}}\right) \ctrlmerge \left(\AllowChoice*{\ell}{E_{2,1}}{E_{2,2}}\right) &\defeq \AllowChoice*{\ell}{E_{1,1} \ctrlmerge E_{1,2}}{E_{2,1} \ctrlmerge E_{2,2}}
  \\[\vertrulegap]
  \left(\CtrlITE*{E_{1,1}}{E_{1,2}}{E_{1,3}}\right) \ctrlmerge \left(\CtrlITE*{E_{2,1}}{E_{2,2}}{E_{2,3}}\right) &\defeq \CtrlITE*{(E_{1,1} \ctrlmerge E_{2,1})}{(E_{1,2} \ctrlmerge E_{2,2})}{(E_{1,3} \ctrlmerge E_{2,3})}
  \\[\vertrulegap]
  \left(\AmIIn*{\rho}{E_{1,1}}{E_{1,2}}\right) \ctrlmerge \left(\AmIIn*{\rho}{E_{2,1}}{E_{2,2}}\right) &\defeq \AmIIn*{\rho}{(E_{1,1} \ctrlmerge E_{2,1})}{(E_{1,2} \ctrlmerge E_{2,2})}
\end{align*}
\endgroup


\subsection{Endpoint Projection}
\label{sec:proj-def}
Note that $\AmI{\ell}{E_1}{E_2}$ is shorthand for $\AmIIn{\{\ell\}}{E_1}{E_2}$.

\begingroup
\allowdisplaybreaks
\rulefiguresize
\begin{align*}
  \epp{X}{L} & = X
  \\[\vertrulegap]
  \epp{\rho.e}{L} & =
    \begin{cases}
      \Ret{e} & \text{if}~L \in \rho \\
      \CtrlUnit & \text{otherwise}
  \end{cases}
  \\[\vertrulegap]
  \epp{\Fun{F}{X \ty \tau}{C}}{L} & = \CtrlFun{F}{X}{\epp{C}{L}}
  \\[\vertrulegap]
  \epp{C_1~C_2}{L} & = \epp{C_1}{L}~\epp{C_2}{L}
  \\[\vertrulegap]
  \epp{\TFun{\alpha \knd \kappa}{C}}{L} & =
  \begin{cases}
    \CtrlTFun{\alpha \knd *_\ell}{\AmI{\alpha}{\epp{\subst{C}{\alpha}{L}}{L}}{\epp{C}{L}}} & \text{if}~ \kappa = *_\ell \\
    \CtrlTFun{\alpha \knd *_s}{\AmIIn{\alpha}{\epp{\subst{C}{\alpha}{\{L\} \cup \alpha}}{L}}{\epp{C}{L}}} & \text{if}~ \kappa = *_s \\
    \CtrlTFun{\alpha \knd \kappa}{\epp{C}{L}} & \text{otherwise}
  \end{cases}
  \\[\vertrulegap]
  \epp{C~t}{L} & = \epp{C}{L}~t
  \\[\vertrulegap]
  \epp{\Fold{C}}{L} & = \CtrlFold{\epp{C}{L}}
  \\[\vertrulegap]
  \epp{\Unfold{C}}{L} & = \CtrlUnfold{\epp{C}{L}}
  \\[\vertrulegap]
  \epp{(C_1,C_2)}{L} & = (\epp{C_1}{L},\epp{C_2}{L})
  \\[\vertrulegap]
  \epp{\Fst{C}}{L} & = \CtrlFst{\epp{C}{L}}
  \\[\vertrulegap]
  \epp{\Snd{C}}{L} & = \CtrlSnd{\epp{C}{L}}
  \\[\vertrulegap]
  \epp{\Inl{C}}{L} & = \CtrlInl{\epp{C}{L}}
  \\[\vertrulegap]
  \epp{\Inr{C}}{L} & = \CtrlInr{\epp{C}{L}}
  \\[\vertrulegap]
  \epp{\Case*{C}{X}{C_1}{Y}{C_2}}{L} & = \CtrlCase*{\epp{C}{L}}{X}{\epp{C_1}{L}}{Y}{\epp{C_2}{L}}
  \\[\vertrulegap]
  \epp{\LetIn{\rho.x \ty t_e}{C_1}{C_2}}{L} & =
  \begin{cases}
    \CtrlLetIn{x}{\epp{C_1}{L}}{\epp{C_2}{L}} & \text{if}~L \in \rho\\
    \epp{C_1}{L} \seqfun \epp{C_2}{L} & \text{if}~L \notin \rho~\text{and}~x \notin \fv{\epp{C_2}{L}}\\
    \CtrlNone & \text{otherwise}
  \end{cases}
  \\[\vertrulegap]
  \epp{\LetIn{\rho.\alpha \knd *_e}{C_1}{C_2}}{L} & =
  \begin{cases}
    \CtrlLetIn{\alpha \knd *_e}{\epp{C_1}{L}}{\epp{C_2}{L}} & \text{if}~L \in \rho\\
    \epp{C_1}{L} \seqfun \epp{C_2}{L} & \text{if}~L \notin \rho~\text{and}~\alpha \notin \fv{\epp{C_2}{L}}\\
    \CtrlNone & \text{otherwise}
  \end{cases}
  \\[\vertrulegap]
  \addtocounter{numlevels}{1}
  \epp{\LetIn{\rho.\alpha \knd *_\ell}{C_1}{C_2}}{L} & =
  \begin{cases}
    \CtrlLetIn*{\alpha \knd *_\ell}{\epp{C_1}{L}}{\AmI{\alpha}{\epp{\subst{C_2}{\alpha}{L}}{L}}{\epp{C_2}{L}}} & \text{if}~L \in \rho\\
    \epp{C_1}{L} \seqfun \epp{C_2}{L} & \text{if}~L \notin \rho ~\text{and}~ \alpha \notin \fv{\epp{C_2}{L}}\\
    \Undef & \text{otherwise}
  \end{cases}
  \\[\vertrulegap]
  \epp{\LetIn{\rho.\alpha \knd *_s}{C_1}{C_2}}{L} & =
  \begin{cases}
    \CtrlLetIn*{\alpha \knd *_s}{\epp{C_1}{L}}{\AmIIn*{\alpha}{\epp{\subst{C_2}{\alpha}{\{L\} \cup \alpha}}{L}}{\epp{C_2}{L}}} & \text{if}~L \in \rho\\
    \epp{C_1}{L} \seqfun \epp{C_2}{L} & \text{if}~L \notin \rho~\text{and}~\alpha \notin \fv{\epp{C_2}{L}}\\
    \CtrlNone & \text{otherwise}
  \end{cases}
  \addtocounter{numlevels}{-1}
  \\[\vertrulegap]
  \epp{C \ChorSend[\ell] \rho}{L} & =
    \begin{cases}
      \SendTo{\epp{C}{L}}{\rho} & \text{if}~L = \ell\\
      \epp{C}{L} \seqfun \RecvFrom{\ell} & \text{if}~L \neq \ell ~\text{and}~ L \in \rho\\
      \epp{C}{L} & \text{otherwise}
    \end{cases}
  \\[\vertrulegap]
  \epp{\syncs{\ell}{d}{\rho} \seq C}{L} & =
    \begin{cases}
      \ChooseFor{d}{\rho}{\epp{C}{L}} & \text{if}~L = \ell\\
      \AllowOneChoice{\ell}{\Left}{\epp{C}{L}} & \text{if}~L \neq \ell ~\text{and}~ L \in \rho ~\text{and}~ d = \Left \\
      \AllowOneChoice{\ell}{\Right}{\epp{C}{L}} & \text{if}~L \neq \ell ~\text{and}~ L \in \rho ~\text{and}~ d = \Left \\
      \epp{C}{L} & \text{otherwise}
  \end{cases}
  \\[\vertrulegap]
  \epp{\ITE{C}{C_1}{C_2}}{L} & =
    \begin{cases}
      \CtrlITE{\epp{C}{L}}{\epp{C_1}{L}}{\epp{C_2}{L}} & \text{if}~L \in \rho\\
      \epp{C}{L} \seqfun \epp{C_1}{L} \sqcup \epp{C_2}{L} & \text{otherwise}
    \end{cases}
\end{align*}
\endgroup

\subsection{Locations Named by a Type or Choreography}
\label{sec:pn-def}
\begingroup
\allowdisplaybreaks
\rulefiguresize
\begin{align*}
  \namedlocs{\alpha} &= \varnothing
  \\[\vertrulegap]
  \namedlocs{L} &= \{L\}
  \\[\vertrulegap]
  \namedlocs{\{\ell\}} &= \namedlocs{\ell}
  \\[\vertrulegap]
  \namedlocs{\rho_1 \cup \rho_2} &= \namedlocs{\rho_1} \cup \namedlocs{\rho_2}
  \\[\vertrulegap]
  \namedlocs{X} &= \varnothing
  \\[\vertrulegap]
  \namedlocs{\rho.e} &= \namedlocs{\rho}
  \\[\vertrulegap]
  \namedlocs{\Fun{F}{X \ty \tau}{C}} &= \namedlocs{C}
  \\[\vertrulegap]
  \namedlocs{C_1~C_2} &= \namedlocs{C_1} \cup \namedlocs{C_2}
  \\[\vertrulegap]
  \namedlocs{\TFun{\alpha \knd \kappa}{C}} &= \namedlocs{C}
  \\[\vertrulegap]
  \namedlocs{C~t} &= \namedlocs{C} \cup \namedlocs{t}
  \\[\vertrulegap]
  \namedlocs{\Fold{C}} &= \namedlocs{C}
  \\[\vertrulegap]
  \namedlocs{\Unfold{C}} &= \namedlocs{C}
  \\[\vertrulegap]
  \namedlocs{(C_1,C_2)} &= \namedlocs{C_1} \cup \namedlocs{C_2}
  \\[\vertrulegap]
  \namedlocs{\Fst{C}} &= \namedlocs{C}
  \\[\vertrulegap]
  \namedlocs{\Snd{C}} &= \namedlocs{C}
  \\[\vertrulegap]
  \namedlocs{\Inl{C}} &= \namedlocs{C}
  \\[\vertrulegap]
  \namedlocs{\Inr{C}} &= \namedlocs{C}
  \\[\vertrulegap]
  \operatorname{LN}\left(\Case*{C}{X}{C_1}{Y}{C_2}\right) &= \namedlocs{C} \cup \namedlocs{C_1} \cup \namedlocs{C_2}
  \\[\vertrulegap]
  \namedlocs{\LetIn{\rho.x \ty t_e}{C_1}{C_2}} &= \namedlocs{\rho} \cup \namedlocs{C_1} \cup \namedlocs{C_2}
  \\[\vertrulegap]
  \namedlocs{\LetIn{\rho.\alpha \knd \kappa}{C_1}{C_2}} &= \namedlocs{\rho} \cup \namedlocs{C_1} \cup \namedlocs{C_2}
  \\[\vertrulegap]
  \namedlocs{C \ChorSend[\ell] \rho} &= \namedlocs{\ell} \cup \namedlocs{\rho} \cup \namedlocs{C}
  \\[\vertrulegap]
  \namedlocs{\syncs{\ell}{d}{\rho} \seq C} &= \namedlocs{\ell} \cup \namedlocs{\rho} \cup \namedlocs{C}
  \\[\vertrulegap]
  \namedlocs{\ITE{C}{C_1}{C_2}} &= \namedlocs{\rho} \cup \namedlocs{C} \cup \namedlocs{C_1} \cup \namedlocs{C_2}
\end{align*}
\endgroup


\subsection{The Less-Than Relation}
\label{sec:less-nonderm-def}
\begin{mathparpagebreakable}
  \infer[]{~}
  {\CtrlNone \lessnd E}
  \and
  \infer{E_1 \lessnd E_2 \\ \CtrlVal{V}}
  {E_1 \lessnd V \CtrlSeq E_2}
  \and
  \infer[]{~}
  {X \lessnd X}
  \and
  \infer[]{~}
  {\CtrlUnit \lessnd \CtrlUnit}
  \and
  \infer[]{~}
  {\Ret{e} \lessnd \Ret{e}}
  \and
  \infer[]{E_{1,1} \lessnd E_{2,1} \\
    E_{1,2} \lessnd E_{2,2}}
  {E_{1,1} \CtrlSeq E_{1,2} \lessnd E_{2,1} \CtrlSeq E_{2,2}}
  \and
  \infer[]{E_1 \lessnd E_2}
  {\CtrlFun{F}{X}{E_1} \lessnd \CtrlFun{F}{X}{E_2}}
  \and
  \infer[]{E_{1,1} \lessnd E_{2,1} \\
  E_{1,2} \lessnd E_{2,2}}
  {E_{1,1}~E_{1,2} \lessnd E_{2,1}~E_{2,2}}
  \and
  \infer[]{E_1 \lessnd E_2}
  {\CtrlTFun{\alpha}{E_1} \lessnd \CtrlTFun{\alpha}{E_2}}
  \and
  \infer[]{E_1 \lessnd E_2}
  {E_1~t \lessnd E_2~t}
  \and
  \infer[]{E_1 \lessnd E_2}
  {\CtrlFold{E_1} \lessnd \CtrlFold{E_2}}
  \and
  \infer[]{E_1 \lessnd E_2}
  {\CtrlUnfold{E_1} \lessnd \CtrlUnfold{E_2}}
  \and
  \infer[]{E_{1,1} \lessnd E_{2,1} \\
    E_{1,2} \lessnd E_{2,2}}
  {(E_{1,1},E_{1,2}) \lessnd (E_{2,1},E_{2,2})}
  \and
  \infer[]{E_1 \lessnd E_2}
  {\CtrlFst{E_1} \lessnd \CtrlFst{E_2}}
  \and
  \infer[]{E_1 \lessnd E_2}
  {\CtrlSnd{E_1} \lessnd \CtrlSnd{E_2}}
  \and
  \infer[]{E_1 \lessnd E_2}
  {\CtrlInl{E_1} \lessnd \CtrlInl{E_2}}
  \and
  \infer[]{E_1 \lessnd E_2}
  {\CtrlInr{E_1} \lessnd \CtrlInr{E_2}}
  \and
  \infer[]{E_{1,1} \lessnd E_{2,1} \\
    E_{1,2} \lessnd E_{2,2} \\
    E_{1,3} \lessnd E_{2,3}}
  {\CtrlCase*{E_{1,1}}{X}{E_{1,2}}{Y}{E_{1,3}} \lessnd \CtrlCase*{E_{2,1}}{X}{E_{2,2}}{Y}{E_{2,3}}}
  \and
  \infer[]{E_{1,1} \lessnd E_{2,1} \\
    E_{1,2} \lessnd E_{2,2}}
  {\CtrlLetIn{x}{E_{1,1}}{E_{1,2}} \lessnd \CtrlLetIn{x}{E_{2,1}}{E_{2,2}}}
  \and
  \infer[]{E_{1,1} \lessnd E_{2,1} \\
    E_{1,2} \lessnd E_{2,2}}
  {\CtrlLetIn{\alpha \knd \kappa}{E_{1,1}}{E_{1,2}} \lessnd \CtrlLetIn{\alpha \knd \kappa}{E_{2,1}}{E_{2,2}}}
  \and
  \infer[]{E_1 \lessnd E_2}
  {\SendTo{E_1}{\rho} \lessnd \SendTo{E_2}{\rho}}
  \and
  \infer[]{~}
  {\RecvFrom{\ell} \lessnd \RecvFrom{\ell}}
  \and
  \infer[]{E_1 \lessnd E_2}
  {\ChooseFor{d}{\ell}{E_1} \lessnd \ChooseFor{d}{\ell}{E_2}}
  \and
  \infer[]{E_{1,1} \lessnd E_{2,1} \\
    E_{1,2} \lessnd E_{2,2}}
  {\AllowChoice*{\ell}{E_{1,1}}{E_{1,2}} \lessnd \AllowChoice*{\ell}{E_{2,1}}{E_{2,2}}}
  \and
  \infer[]{E_{1,1} \lessnd E_{2,1} \\
    E_{1,2} \lessnd E_{2,2} \\
    E_{1,3} \lessnd E_{2,3}}
  {\CtrlITE*{E_{1,1}}{E_{1,2}}{E_{1,3}} \lessnd \CtrlITE*{E_{2,1}}{E_{2,2}}{E_{2,3}}}
  \and
  \infer[]{E_{1,1} \lessnd E_{2,1} \\
    E_{1,2} \lessnd E_{2,2}}
  {\AmIIn*{\rho}{E_{1,1}}{E_{1,2}} \lessnd \AmIIn*{\rho}{E_{2,1}}{E_{2,2}}}
\end{mathparpagebreakable}


\end{document}